\pdfoutput=1
\RequirePackage{fix-cm}
\pdfoutput=1
\documentclass[smallcondensed]{svjour3}     
\smartqed  

\usepackage{graphicx}
\usepackage{multirow}
\usepackage{amsmath}
\usepackage{subfig}
\usepackage{wrapfig,lipsum,booktabs}
\usepackage{cite}
\usepackage{url}
\usepackage{longtable}
\usepackage{natbib}
\usepackage{lscape}
\setcitestyle{aysep={}} 
\usepackage{afterpage}
\usepackage{array}
\usepackage{tabu}
\usepackage[title,toc,page]{appendix}
\usepackage{xcolor}
\usepackage{pifont}
\usepackage{amssymb}

\begin{document}

\title{Inferring User Interests in Microblogging Social Networks: A Survey}


\author{Guangyuan Piao         \and
        John G. Breslin 
}


\institute{Guangyuan Piao \at
              Insight Centre for Data Analytics, Data Science Institute, National University of Ireland Galway \\
              Tel.: +353-83-4519532\\
              \email{guangyuan.piao@insight-centre.org}           
           \and
           John G. Breslin \at
            Insight Centre for Data Analytics, Data Science Institute, National University of Ireland Galway \\
            Tel.: +353-91-492622\\
}

\date{Received: date / Accepted: date}

\maketitle

\begin{abstract}
With the growing popularity of microblogging services such as Twitter in recent years, an increasing number of users are using these services in their daily lives. The huge volume of information generated by users raises new opportunities in various applications and areas. Inferring user interests plays a significant role in providing personalized recommendations on microblogging services, and also on third-party applications providing social logins via these services, especially in cold-start situations. In this survey, we review user modeling strategies with respect to inferring user interests from previous studies. To this end, we focus on four dimensions of inferring user interest profiles: (1) \emph{data collection}, (2) \emph{representation} of user interest profiles, (3) \emph{construction and enhancement} of user interest profiles, and (4) the \emph{evaluation} of the constructed profiles. Through this survey, we aim to provide an overview of state-of-the-art user modeling strategies for inferring user interest profiles on microblogging social networks with respect to the four dimensions. For each dimension, we review and summarize previous studies based on specified criteria. Finally, we discuss some challenges and opportunities for future work in this research domain.

\keywords{User Modeling \and User Interests \and User Profiles \and Social Web \and Microblogging \and Twitter \and Social Networks \and Information Filtering \and Recommender Systems \and Personalization \and Survey}
\end{abstract}



\sloppy
\section{Introduction}
\label{intro}


Microblogging\footnote{\url{https://en.wikipedia.org/wiki/Microblogging}} social networks such as Twitter\footnote{\url{https://twitter.com/}} and Facebook\footnote{\url{https://www.facebook.com/}} are being widely used in our daily lives. Twitter and Facebook have 328 million and 2 billion monthly active users\footnote{\url{https://www.omnicoreagency.com/twitter-statistics/}}\footnote{\url{https://www.omnicoreagency.com/facebook-statistics/}}, which shows the popularity of these services. The abundant information generated by users in OSNs creates new opportunities for inferring user interest profiles, which can be used for providing personalized recommendations to those users either on those OSNs or on third-party services allowing social login functionality\footnote{\url{https://en.wikipedia.org/wiki/Social_login}} from the same OSNs. Social login is a technology which allows visitors to a website to log in using their OSN accounts rather than having to register a new one\footnote{\url{https://hbr.org/2011/10/social-login-offers-new-roi-fr}}. A recent survey showed that over 94\% of 18-34 year olds have used social login via Twitter, Facebook, etc.\footnote{\url{http://www.gigya.com/blog/why-millennials-demand-social-login/}} With the continued widespread development of the social login functionality, inferring user interest profiles from their OSN activities plays a central role in many applications for providing personalized recommendations with the permission of those users, especially for cold-start users who have joined those services recently. 

In the literature, there have been many studies that focused on inferring user interest profiles with different purposes such as providing personalized recommendations with respect to news \citep{Abel2011g, Gao:2011:ITU:2052138.2052335}, research articles \citep{Bolting2015, Nishioka:2016:PVT:2910896.2910898}, and Points Of Interest (POI) \citep{Abel2012a}. Despite the popularity of inferring user interests in OSNs, there is a lack of an extensive review on user modeling strategies for inferring user interest profiles in OSNs. To our knowledge, only one related short survey \citep{Abdel-Hafez2013} has been formally published. \cite{Abdel-Hafez2013} provided a general overview of user modeling in social media websites which includes all types of OSNs without focusing on a specific type. As a result, the details of user modeling techniques for microblogging websites were not presented in \cite{Abdel-Hafez2013}. For example, including OSNs such as Delicious\footnote{\url{https://del.icio.us/}} and Flickr\footnote{\url{https://www.flickr.com/}} which  are based on \emph{folksonomies} (folks taxonomies) together with microblogging OSNs for a single survey presents some difficulties due to the volume of literature on \emph{folksonomy}-based user modeling \citep[e.g.,][to name a few]{Hung2008, Szomszor2008, Abel2011c, Mezghani:2012:UPM:2187980.2188230, Carmagnola2008a}. In addition, the survey conducted by \cite{Abdel-Hafez2013} does not cover studies from recent years. In this survey, we focus in particular on user modeling strategies in microblogging OSNs in terms of several user modeling dimensions, and analyze over 50 studies including more recent ones (see Appendix \ref{appendix:works} for details of the surveyed studies).


There has been a varied set of terms used to denote inferring user interests in the literature, such as ``user (interest) modeling/profiling/detection'', ``inferring/modeling/predicting user interests''. User modeling/profiling, as a broad term, may refer to different meanings without a specific definition. A general definition of \textit{user profiling} given by \cite{Zhou2012} is ``the process of acquiring, extracting and representing the features of users''. Similarly, in \cite{Brusilovsky2007}, the \emph{user model} is defined in the context of adaptive systems as ``a representation of information about an individual user that is essential for an adaptive system to provide the adaptation effect''. Based on a specific definition of what the \emph{features} and \emph{information} are in these definitions by \cite{Zhou2012} and \cite{Brusilovsky2007}, the corresponding user models/profiles and the process of obtaining them might be different.

\begin{figure*}[!b]
\includegraphics[width=\textwidth]{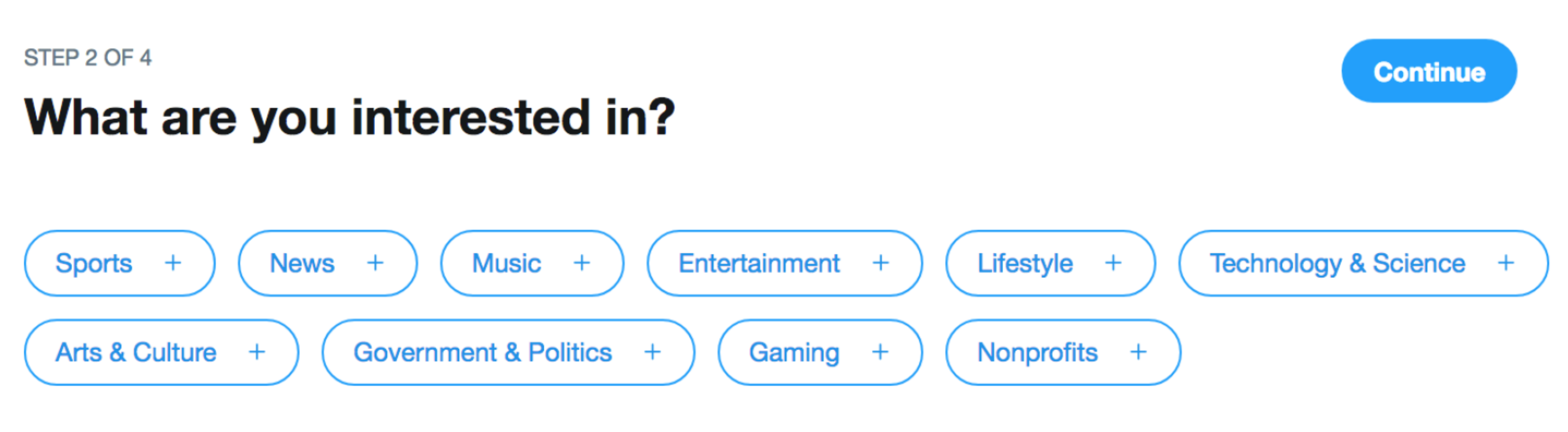}
\caption{Explicit information from users during signing up on Twitter.}
\label{explicit}       
\end{figure*}

\cite{Rich1979} along with \cite{Cohen1979} and \cite{Perrault1978}, where the terms \emph{user model} and \emph{user modeling} can be traced back to, also pointed out the need for classifying your user model as it might refer to several different things without a proper definition. Three major dimensions were used in \cite{Rich1979} for classifying user models: 

\begin{itemize}
	\item Are they models of a canonical user or are they models of individual users?
	\item Are they constructed explicitly by the user themselves or are they abstracted by the system on the basis of the user's behavior?
	\item Do they contain short-term or long-term information?
\end{itemize}
Explicit information denotes the information which requires direct input by users such as surveys or forms, which will impose an additional burden on the users. Figure \ref{explicit} shows an example of collecting \emph{explicit} information about user interests during sign up on Twitter for the first time. 

\subsection{Definition of User Modeling in This Survey}

In the context of research on inferring user interests on OSNs, most studies have focused on exploiting \emph{implicit} information such as the posts of users in order to infer user interest profiles. Based on the classification criteria from \cite{Rich1979}, user models discussed in this survey are about individual users constructed implicitly based on their activities. For the third criterion used in \cite{Rich1979}, there is no clear cut option as both short- and long-term information have been used in different user modeling strategies in the literature. In addition, user models can refer to various types of information relevant for each user in the domain of OSNs. For example, they might contain basic information such as age, gender, country, etc., or keywords that represent their interests. In this paper, we focus particularly on user models with respect to user interests. Although several terms such as ``user model" and ``user profile" have been used interchangeably in the literature, here we formally define these terms as follows:

\begin{definition}[User Model]
A \emph{user model} is a (data) structure that is used to capture certain \emph{characteristics} about an individual user, and a \emph{user profile} is the actual representation in a given user model. The process of obtaining the user profile is called \emph{user modeling}.
\end{definition}




Given this definition of a user model and the classification criteria from \cite{Rich1979}, user model in this survey aims to capture user \emph{interests} with respect to an \emph{individual} user \emph{implicitly} based on \emph{long-term} or \emph{short-term} knowledge via a user modeling strategy, to derive the interest profile of that user. 

\begin{figure*}[!h]
\includegraphics[width=\textwidth]{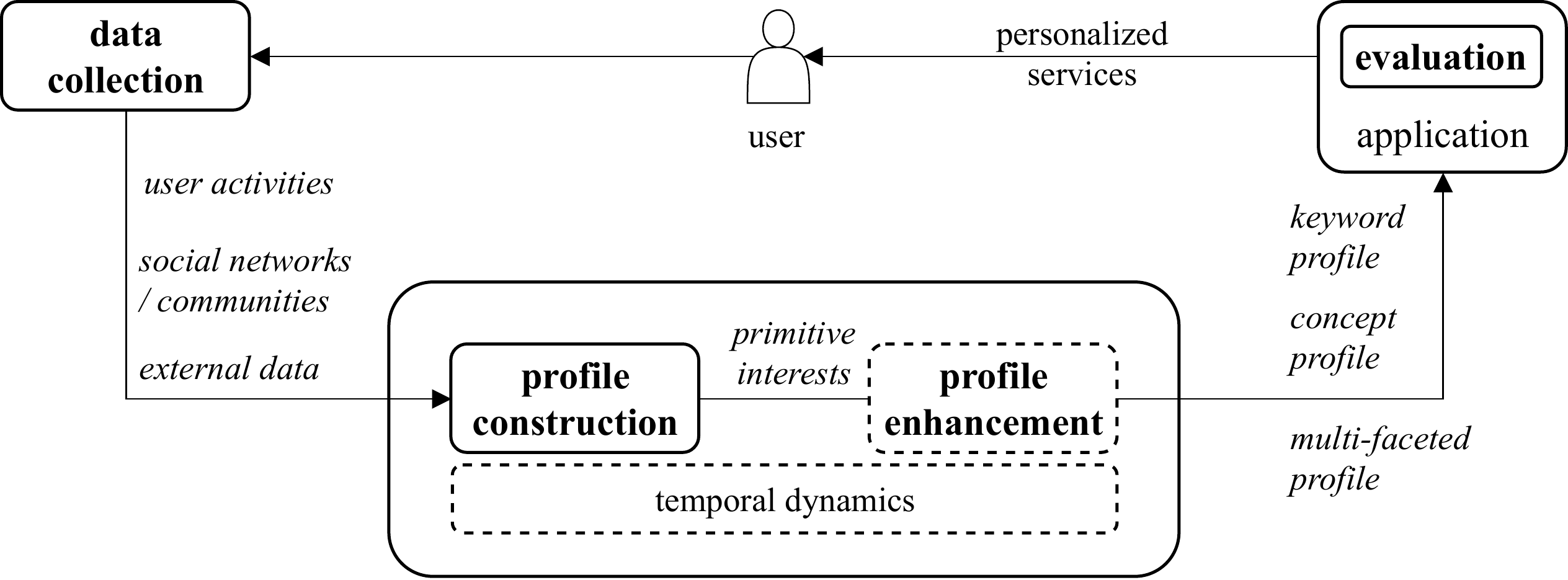}
\caption{Overview of user profile-based personalization process}
\label{fig:2}       
\end{figure*}

Figure \ref{fig:2} presents an overview of the modified user profile-based personalization process from \cite{Abdel-Hafez2013} and \cite{Gauch:2007:UPP:1768197.1768200}. We modified the process from \cite{Abdel-Hafez2013} in order to reflect different aspects of user modeling strategies proposed in previous studies in the context of OSNs in detail. For example, we focus on data collection from \emph{user activities}, \emph{social networks/communities} or \emph{external data} of an OSN instead of \emph{explicit} or \emph{implicit} feedback as most previous studies have focused on exploiting \emph{implicit} information for inferring user interests. The modified user profile-based personalization process consists of three main phases. The first step is collecting data which will be used for inferring user interests. Subsequently, user interest profiles are constructed based on the data collected. We use \emph{primitive interests} \citep{Kapanipathi2014} to denote the interests directly extracted from the collected data. Those primitive interests can either be used as the final output of a profile constructor or can be further enhanced, e.g., based on background knowledge from Knowledge Bases (KBs) such as Wikipedia\footnote{\url{www.wikipedia.org}}. The output of the profile constructor is user interest profiles represented based on a predefined representation of interest profiles, e.g., word-based user interest profiles. Finally, the constructed user profiles are evaluated, and can be used in specific applications such as recommender systems for personalized recommendations. 

In this paper, we mainly discuss four dimensions of the user modeling process: (1) \emph{data collection}, (2) \emph{representation} of user interest profiles, (3) \emph{profile construction and enhancement}, and (4) the \emph{evaluation} of the constructed user interest profiles. In summary, the contribution of this paper is threefold. 

\begin{itemize}
	\item First, we provide a detailed review of user modeling approaches on microblogging services in terms of the three phases in Figure \ref{fig:2} with the following focuses:
	 	\begin{enumerate}
			\item \emph{What information is used for inferring user interest profiles?}
			\item \emph{How are the user interest profiles represented?}
			\item \emph{How are the user interest profiles constructed?}
			\item \emph{How are the constructed user profiles evaluated?}
		\end{enumerate}
	\item Second, we summarize the approaches with respect to these focuses based on specified criteria to be specified later on.
	\item Finally, we discuss the challenges and opportunities based on the strengths and weaknesses of different approaches.
\end{itemize}

\renewcommand{\arraystretch}{1.2}
\begin{table}[!b]
\centering
\caption{Online Social Networks used for previous studies.}
\label{osns}
\begin{tabular}{|l|l|}
\hline
\begin{tabular}[c]{@{}l@{}}  \textbf{OSNs} \\ \textbf{(\# of studies)} \end{tabular} & \textbf{Examples} \\ \hline
Twitter (47)                                                                      & \begin{tabular}[c]{@{}l@{}} \cite{Chen2010}, \cite{Lu2012}, \cite{Kapanipathi2014, Kapanipathi2011}, 
\\ \cite{piao2016exploring, Guangyuan2017, Piao2017, Piao2016b, Piao2016d}, \cite{Besel:2016:ISI:2851613.2851819, Besel:2016:QSI:3015297.3015298}, 
\\ \cite{Abel2011g, Abel2011e, Abel2012a,Abel2011d, Abel:2013:TUM:2540128.2540558}, \cite{Siehndel:2012:TUP:2887379.2887395}, 
\\ \cite{Michelson2010}, \cite{Bhattacharya:2014:IUI:2645710.2645765}, \\ \cite{Orlandi2012}, \cite{Hannon2012}, \cite{Jiang2015}, 
\\ \cite{Budak2014}, \cite{Faralli2015, Faralli2017}, \cite{Weng:2010:TFT:1718487.1718520},
\\ \cite{Zarrinkalam2015a, Zarrinkalam2016}, 
\\ \cite{Narducci2013}, \cite{Xu2011}, \cite{GarciaEsparza:2013:CCT:2449396.2449402}, 
\\ \cite{Nishioka:2016:PVT:2910896.2910898,Nishioka:2015:ITU:2809563.2809601}, \cite{Gao:2011:ITU:2052138.2052335}, 
\\ \cite{Vu:2013:IMU:2505515.2507883}, \cite{Phelan:2009:UTR:1639714.1639794}, \cite{Penas2013},
\\ \cite{Sang:2015:PFT:2806416.2806470}, \cite{Karatay2015a}, 
\\ \cite{Kanta2012}, \cite{OBanion2012}, \cite{Nechaev},
\\ \cite{Lim:2013:ICT:2491055.2491078}, \cite{Bolting2015}, 
\\ \cite{AnilKumarTrikhaFattaneZarrinkalam}, \cite{Spasojevic:2014:LLS:2623330.2623350}, \cite{Jipmo2017}
\end{tabular}     \\ \hline
Facebook (7)                                                                     & \begin{tabular}[c]{@{}l@{}} \cite{Kang2016},  \cite{Orlandi2012},  \cite{Kapanipathi2011}, \\ \cite{Narducci2013}, \cite{Bhargava:2015:UMU:2678025.2701365}, \cite{Ahn:2012:IUI:2457524.2457681}, \\\cite{Spasojevic:2014:LLS:2623330.2623350} \end{tabular}  \\ \hline
LinkedIn (2)                                                             & \cite{Kapanipathi2011}, \cite{Spasojevic:2014:LLS:2623330.2623350}       \\ \hline
Google+\footnotemark (1)                                                                & \cite{Spasojevic:2014:LLS:2623330.2623350}       \\ \hline
\end{tabular}
\end{table}
\footnotetext{\url{https://plus.google.com/}}

Table \ref{osns} provides a summary of OSNs used for the works discussed in this survey. As we can see from the table, Twitter has been widely used due to its popularity and the higher degree of openness. Other OSNs such as Facebook or LinkedIn\footnote{\url{https://www.linkedin.com/}} need to gain the permissions of users to access their data. Therefore, users have to be recruited for conducting an experiment, which results in less studies using these OSNs. In contrast to other studies, the study from Klout\footnote{\url{https://klout.com/}}, Inc. \citep{Spasojevic:2014:LLS:2623330.2623350}, which is a social media platform that aggregates and analyzes data from multiple OSNs, leveraged all the OSNs listed in Table \ref{osns}. As different design choices can be made for user modeling with different purposes, Table \ref{purpose} provides an overview of the purpose of user modeling in each study. As we can see from the table, the majority of the previous studies have been conducted with the purpose of predicting user interests followed by recommending different types of content such as news, URLs, publications, and tweets.

\begin{table}[!h]
\centering
\caption{Purposes of user modeling in OSNs from previous studies.}
\label{purpose}
\begin{tabular}{|l|l|}
\hline
\textbf{\begin{tabular}[c]{@{}l@{}}Purpose\end{tabular}} 
 & \textbf{Examples} \\ \hline
\begin{tabular}[c]{@{}l@{}} Predicting \\ user interests      \end{tabular}                                                                                                            & \begin{tabular}[c]{@{}l@{}} \cite{Kapanipathi2014}, \cite{Kang2016},  \\  \cite{Michelson2010}, \cite{Budak2014}, \\ \cite{Bhattacharya:2014:IUI:2645710.2645765}, \cite{Besel:2016:ISI:2851613.2851819, Besel:2016:QSI:3015297.3015298}, \\ \cite{Orlandi2012}, \cite{Narducci2013}, \\ \cite{Bhargava:2015:UMU:2678025.2701365}, \cite{GarciaEsparza:2013:CCT:2449396.2449402}, \\ \cite{Vu:2013:IMU:2505515.2507883}, \cite{Ahn:2012:IUI:2457524.2457681}, \cite{Abel2011e}
\\ \cite{Zarrinkalam2016}, \cite{Ahn:2012:IUI:2457524.2457681}, \\ \cite{Spasojevic:2014:LLS:2623330.2623350}, \cite{Jipmo2017},
\\ \cite{Faralli2017}, \cite{Jiang2015}, \\ \cite{Xu2011}, \cite{Penas2013}, \cite{Lim:2013:ICT:2491055.2491078}   \end{tabular}    \\ \hline
\begin{tabular}[c]{@{}l@{}} News \\recommendations \end{tabular}  &  \begin{tabular}[c]{@{}l@{}} \cite{Abel2011g}, \cite{Gao:2011:ITU:2052138.2052335}, \\ \cite{Zarrinkalam2015a}, \cite{Sang:2015:PFT:2806416.2806470},  \\ \cite{Kanta2012}, \cite{OBanion2012}  \end{tabular} \\ \hline
\begin{tabular}[c]{@{}l@{}} URL \\recommendations    \end{tabular}                                                                                                       & \begin{tabular}[c]{@{}l@{}} \cite{Chen2010}, \cite{Abel2011d}, \\ \cite{piao2016exploring, Piao2016a, Guangyuan2017, Piao2017, Piao2016b, Piao2016d}   \end{tabular}   \\ \hline
\begin{tabular}[c]{@{}l@{}} Publication \\recommendations    \end{tabular}                                                                                                       & \cite{Nishioka:2016:PVT:2910896.2910898},  \cite{Bolting2015}        \\ \hline
\begin{tabular}[c]{@{}l@{}} Tweet \\recommendations    \end{tabular}                                                                                                                & \begin{tabular}[c]{@{}l@{}} \cite{Lu2012}, \cite{Sang:2015:PFT:2806416.2806470}, \\ \cite{Karatay2015a}, \cite{AnilKumarTrikhaFattaneZarrinkalam}  \end{tabular}         \\ \hline
\begin{tabular}[c]{@{}l@{}}  Researcher \\recommendations        \end{tabular}                                                                                                             & \cite{Nishioka:2015:ITU:2809563.2809601}         \\ \hline
POI recommendations &  \cite{Abel2012a} \\ \hline
\begin{tabular}[c]{@{}l@{}} User recommendations \\ and classifications   \end{tabular} &  \cite{Faralli2015} \\ \hline
\begin{tabular}[c]{@{}l@{}} Concealing \\ user interests   \end{tabular} &  \cite{Nechaev} \\ \hline
\end{tabular}
\end{table}

\renewcommand{\arraystretch}{1.2}
\begin{table}[!t]
\centering
\caption{Conceptual framework for discussing the related work in this survey.}
\label{conceptual_framework}
\begin{tabular}{|p{10cm}|}
\hline
\textbf{Data Collection}  \\ \hline
\begin{enumerate}
	\item using user activities
	\item using the social networks/communities of a user
	\item using external data
\end{enumerate}
\\ \hline
\textbf{Representation of User Interest Profiles} \\ \hline
\begin{enumerate}
	\item keyword profiles
	\item concept profiles
	\item multi-faceted profiles
\end{enumerate}
\\ \hline
\textbf{Construction and Enhancement of User Interest Profiles} \\ \hline
\begin{enumerate}
	\item profile construction with weighting schemes
	\begin{itemize}
		\item heuristic approaches
		\item probabilistic approaches
	\end{itemize}
	\bigskip
	\item profile enhancement
	\begin{itemize}
		\item leveraging hierarchical knowledge
		\item leveraging graph-based knowledge
		\item leveraging collective knowledge
	\end{itemize}
	\bigskip
	\item temporal dynamics
	\begin{itemize}
		\item constraint-based approaches
		\item interest decay functions
	\end{itemize}
\end{enumerate}
\\ \hline
\textbf{Evaluation} \\ \hline
\end{tabular}
\end{table}

{\color{black}Table \ref{conceptual_framework} is a conceptual framework for discussing user modeling strategies proposed in the related work and to act as a ``guide'' to the rest of this survey.}
The rest of this paper is organized as follows. In Section 2, we discuss what kind of information has been collected for inferring user interests. Section 3 introduces various representations of user interest profiles proposed in the literature. In Section 4, we review how user profiles have been constructed based on different dimensions such as considering the temporal dynamics of user interests. In Section 5, we discuss how those constructed user profiles have been evaluated in the literature. Finally, we conclude the paper with some discussions of opportunities and challenges with respect to user modeling on microblogging OSNs in Section 6.


\section{Data Collection}

\subsection{Overview}

This section of the survey discusses the first stage of user modeling, which is the data collection. In the context of OSNs, there are various information sources for collecting data in order to infer user interest profiles such as user information including the tweets or profiles with respect to a user and information from that user's social network. The information used for user modeling is important as it might directly affect later stages such as the representation and construction of user interest profiles, and the quality of final profiles. The discussion is carried out over the criteria of whether the information is collected from a \emph{user's activities} or the \emph{social networks/communities} of that user from the target microblogging platform (where the target users come from) or \emph{external data}. Given Twitter is the largest microblogging social networking platform and is the most used OSNs in the literature as depicted in Table \ref{osns}, here we mainly focus on inferring user interest profiles on Twitter.

\subsubsection{Using user activities}



A straightforward way of inferring user interests for a target user is leveraging information from the user's activities in OSNs. Take Twitter as an example, a user can have different activities such as posting, re-tweeting, liking or replying to a tweet. Users can also describe themselves in their profiles or follow other people on Twitter which might reveal their interests. Therefore, we can leverage these user activities to infer user interests. This could be analyzing data from the posts, profiles or following activities of users. For instance, we can assume that a user is interested in \texttt{Microsoft} if the user mentions \texttt{Microsoft} frequently in the tweets or is following the Twitter account \texttt{@Microsoft}. However, inferring user interests from their activities such as posting tweets or re-tweeting requires users to be active, which is not always the case. For example, \cite{Gong2015} reported that a significant portion of Twitter users are \emph{passive ones} who keep following other users in order to consume information on Twitter but who do not generate any content. 

\subsubsection{Using the social networks/communities of a user}

Leveraging information from the social networks/communities of a user can be useful to infer user interest profiles, especially for \emph{passive users} who have little activity but who keep following other users to receive information. In this case, the generated content such as the posts and the profiles of users in a user's social network can be used for inferring that user's interests. For example, if many followees of a user post tweets with respect to \texttt{Microsoft} frequently or belong to a common community related to \texttt{Microsoft}, we can assume that the user is interested in \texttt{Microsoft} as well. 


\subsubsection{Using external data}

The ideal length of a post on any OSN ranges between 60 to 140 characters for better user engagement\footnote{\url{https://goo.gl/j97H1R}}. Analyzing microblogging services such as Twitter is challenging due to their nature of generating short, noisy texts. Understanding those short messages plays a key role in user modeling in microblogging services. To this end, previous studies have investigated leveraging external data such as the content of embedded links/URLs in a tweet, in order to enrich the short text for a better understanding of it. \cite{Haewoon2010a} showed that most of the topics on Twitter are about news which could also be found in mainstream news sites. In this regard, some researchers have proposed linking microblogs to news articles and exploring the content of news articles in order to understand short texts in microblogging services better. 

\subsection{Review}

\subsubsection{Using user activities}
\label{Using information inside the platform}

The posts generated by users are the most common source of information for inferring user interests. Take Twitter as an example, the tweets or retweets of users provide a great amount of data that might implicitly indicate what kinds of topics a user might be interested in. Therefore, using the post streams of target users for inferring user interest profiles has been widely studied in the literature regardless of the different manners for how user interests are represented.  For instance, \cite{Kapanipathi2014} extracted Wikipedia entities from the tweet streams of users while \cite{Chen2010} extracted keywords from them. Inferring user interests based on users' posts requires users to be active, i.e., continuously generating content. On the one hand, there is an increasing number of users leveraging OSNs to seek the information they need, e.g., one in three Web users look for medical information, and over half of surveyed users consume news in OSNs\footnote{\url{http://bit.ly/pewsnsnews}} \citep{Sheth2016}. On the other hand, there is also a rise of passive users in OSNs. For example, two out of five Facebook users only browse information without active participation within the platform\footnote{\url{http://www.corporate-eye.com/main/facebooks-growing-problem-passive-users/}} \citep{Besel:2016:ISI:2851613.2851819}, and \cite{Gong2015} reported that a significant portion of Twitter users are \emph{passive ones} who consume information on Twitter without generating any content. Therefore, it is also important to infer user interest profiles for those \emph{passive users} in OSNs.

Some studies pointed out that exploring posts for inferring user interests is computationally ineffective and unstable due to the changing interests of users \citep{Besel:2016:QSI:3015297.3015298, Faralli2015, Faralli2017, Besel:2016:ISI:2851613.2851819, Nechaev}. Instead of analyzing posts to infer user interests, these studies proposed using the \emph{followeeship} information of users, which can infer more stable user interest profiles as the relationships of common users tend to be stable \citep{Myers2014}. In this line of work, \emph{topical followees} that can be mapped to Wikipedia entities often need to be identified, e.g., identifying the followee account \texttt{@messi10stats} on Twitter as \texttt{wiki\footnote{The prefix wiki denotes https://en.wikipedia.org/wiki/}:Lionel\_Messi}. One of the problems with these approaches based on topical followees is that only a small portion of users' followees are topical ones. The authors from \cite{Faralli2015} and \cite{Guangyuan2017} both showed that, on average, only 12.7\% and 10\% of followees of users in their datasets can be linked to Wikipedia entities. Therefore, a lot of information from followees that do not have corresponding Wikipedia entities is missed. For example, based on the topical-followees approach we cannot infer any interests for a user who is following \texttt{@Alice} who has a biography as \emph{``User Modeling and Recommender Systems researcher''}.
\\\\
\textbf{Pros and cons.} Analyzing user activities for inferring user interests collects data from users themselves which can reflect their interests better compared to inferring from their social networks which will be discussed later. However, it requires users actively generate content in order to infer their interests from their generated content such as tweets, retweets, and likes on Twitter. Although leveraging the \emph{topical-followees} approach can be used for inferring user interests for passive users, the usage of followees' information is limited.


\subsubsection{Using the social networks/communities of a user}

To cope with some problems such as inferring user interest profiles for passive users, information from social networks such as tweets from followees or followers or posts from Facebook friends can be utilized for inferring user interests for \emph{passive users} as well as \emph{active ones}. All aforementioned activities used for inferring a user's interests can be analyzed with respect to a user's social network as well for inferring that user's interests. For instance, \cite{Chen2010} and \cite{Budak2014} explored the tweets of target users and their followees to infer user interests. Although using posts generated by users is of great potential for mining user interests, it also faces some challenges due to the short and noisy nature of microblogs. Compared to the aforementioned topical-followees approach, information from the social networks of users such as their followees can provide much more information. Returning to the example of inferring user interests for a user who is following \texttt{@Alice} in the previous subsection, we can infer this user is interested in \texttt{User Modeling} and \texttt{Recommender Systems} based on the biography of \texttt{@Alice} - \emph{``User Modeling and Recommender Systems researcher''}. In \cite{Guangyuan2017}, the authors proposed leveraging \emph{biographies} of followees to extract entities instead of mapping followees to Wikipedia entities, and showed the improvement of inferred user interest profiles in the context of URL recommendations. 

\emph{List membership}, which is a kind of ``tagging'' feature on Twitter, has been explored as well. A list membership is a topical list or community which can be generated by any user on Twitter, and the creator of the list can freely add other users to the topical list. For instance, a user \texttt{@Bob} might create a topical list named ``Java'' and add his followees who have been frequently tweeting about news on this topic. Therefore, if a user \texttt{@Alice} is following users who have been added into many topical lists related to the topic \texttt{Java}, it might suggest that \texttt{@Alice} is interested in this topic as well. \cite{Kim2010a} studied the usage of Twitter lists and confirmed that lists can serve as good groupings of Twitter users with respect to their characteristics based on a user study. Based on the study, the authors also suggested that the Twitter list can be a valuable information source in many application domains including recommendations. In this regard, several studies have exploited list memberships of followees to infer user interest profiles \citep{Hannon2012, Bhattacharya:2014:IUI:2645710.2645765, Piao2017}.

User interests might be following global trends in some trends-aware applications such as news recommendations. To investigate it, \cite{Gao:2011:ITU:2052138.2052335} proposed interweaving global trends and personal user interests for user modeling. In addition to leveraging the tweets of a target user for inferring user interests, the authors constructed a trend profile based on all tweets in the dataset in a certain time period. Afterwards, the final user interest profile was built by combining the two profiles. The results showed that combined user interest profiles can improve the performance of news recommendations while the first profile based on personal tweets plays a more significant role in the combination.
\\\\
\textbf{Pros and cons.} On the one hand, a lot of data can be collected from the social networks of users, which is useful in the case of when inferring user interest profiles for passive users who do not generate much content but who keep following other users. On the other hand, it is difficult to distinguish the activities of a user's followees that are relevant to the interests of that user. For example, the followees of a user can tweet a wide range of topics that they are interested in, and the user is not always interested in all those topics.

\subsubsection{Using external data}

One of the challenges of inferring user interests from OSNs is that the generated content is often short and noisy \citep{Bontcheva2014}. To better understand the short texts of microblogging services such as tweets, external information beyond the target platform has been explored on top of the information sources discussed in the previous subsections. For instance, \cite{Abel2011g, Abel2011e, Abel:2013:TUM:2540128.2540558} proposed linking tweets to news articles and extract the \emph{primitive interests} of users based on their tweets as well as the content of related news articles. Several strategies were proposed in \cite{Abel2011e}, which were later on developed as a Twitter-based User Modeling Service \citep[TUMS, ][]{Tao2012}. However, it requires maintaining up-to-date news streams from mainstream news providers such as CNN\footnote{\url{http://edition.cnn.com/}} in order to link tweets to relevant news articles. Instead, \cite{Abel2011d} and \cite{Piao2016d} leveraged the content of the embedded URLs in tweets. \cite{Hannon2012} used a third-party service Listorious\footnote{\url{http://listorious.com}, not available at the time of writing.}, which is a service providing annotated tags of list memberships on Twitter, for inferring user interest profiles. Given a target user \emph{u}, the authors construct \emph{u}'s interest profile based on the tags of list memberships with respect to the user. 


With the popularity of different OSNs, users nowadays tend to have multiple OSN accounts across various platforms \citep{Liu2013b}. In this context, some of the previous studies have investigated exploiting user interest profiles from other OSNs for cross-system user modeling. For instance, \cite{Orlandi2012} and \cite{Kapanipathi2011} presented user modeling applications that can aggregate different user interest profiles from various OSNs. However, the evaluation of aggregated user interest profiles has not been provided. \cite{Abel2012a} investigated cross-system user modeling with respect to POI, and showed that the aggregation of Twitter and Flickr user data yields the best performance in terms of POI recommendations compared to modeling users separately based on a single platform. The result is in line with another study by them which aggregated user interest profiles on social tagging systems such as Delicious\footnote{\url{https://www.delicious.com}}, StumbleUpon\footnote{\url{https://www.stumbleupon.com}}, and Flickr  \citep{Abel2013}. 


The work from Klout \citep{Spasojevic:2014:LLS:2623330.2623350}, which allows their users to add multiple OSN identities on their services, showed many insights on aggregating user information from multiple information sources in different OSNs for inferring user interests. The authors pointed out that using user-generated content (UGC) alone leads to a high precision but low recall for topic recommendations, and therefore, other information sources such as the ones from followees are needed. They also observed that the overlap of a user's interests from different OSNs is very small, which shows that a user may not reveal all his/her interests on any single OSN alone due to the different characteristics of OSNs. Therefore, aggregating users' information in different OSNs leads to a better understanding of their interests \citep{Spasojevic:2014:LLS:2623330.2623350}.
\\\\
\textbf{Pros and cons.} Leveraging external data such as the content of embedded URLs in a tweet can provide a better understanding of short microblogs, and exploring information from other OSNs of users can reveal their interests better compared to exploring a single OSN. Nevertheless, analyzing external data requires an additional effort and it is not always available. In addition, external data can also have irrelevant content with respect to user interests and might introduce some noise.

\subsection{Summary and discussion}
In this section, we reviewed different information sources that have been used for collecting data in order to infer user interest profiles. Table \ref{datacollection} summarizes information sources used for inferring user interest profiles in the literature. As we can see from Table \ref{datacollection}, user activities have been used widely for inferring user interest profiles in microblogging social networks in previous studies.

\renewcommand{\arraystretch}{1.5}
\afterpage{
\begin{landscape}
\begin{table}[]
\centering
\caption{Information used for collecting data for inferring user interest profiles.}
\label{datacollection}
\begin{tabular}{|c|c|c|l|}
\hline
\textbf{\begin{tabular}[c]{@{}l@{}}User \\Activities \end{tabular}} & \textbf{\begin{tabular}[c]{@{}l@{}}Social Networks/\\Communities \end{tabular}} & \textbf{\begin{tabular}[c]{@{}l@{}}External \\Data \end{tabular}}  & \textbf{Examples} \\ \hline

\checkmark            &  & & \begin{tabular}[c]{@{}l@{}} \cite{Lu2012},  \cite{Kapanipathi2014}, \cite{Kang2016},  \\ \cite{piao2016exploring, Piao2016b}, \cite{Orlandi2012}, \cite{Weng:2010:TFT:1718487.1718520}, \\ \cite{Michelson2010}, \cite{Siehndel:2012:TUP:2887379.2887395},  \\ \cite{Zarrinkalam2015a, Zarrinkalam2016}, 
\\ \cite{Jiang2015},  \cite{Narducci2013}, \cite{Xu2011}, 
\\ \cite{Nishioka:2016:PVT:2910896.2910898, Nishioka:2015:ITU:2809563.2809601}, 
\\ \cite{Penas2013}, \cite{Sang:2015:PFT:2806416.2806470}, \cite{OBanion2012},
\\ \cite{Bolting2015}, \cite{Penas2013}, \cite{Jipmo2017}, 
\\ \cite{AnilKumarTrikhaFattaneZarrinkalam}, \cite{Bhargava:2015:UMU:2678025.2701365}, \cite{Ahn:2012:IUI:2457524.2457681},
\\ \cite{Besel:2016:ISI:2851613.2851819, Besel:2016:QSI:3015297.3015298}, \cite{Faralli2015, Faralli2017}, 
\\ \cite{Lim:2013:ICT:2491055.2491078}, \cite{Nechaev}, \cite{Vu:2013:IMU:2505515.2507883} \end{tabular}      \\ \hline

& \checkmark	&		&	\begin{tabular}[c]{@{}l@{}}  \cite{Phelan:2009:UTR:1639714.1639794}, \cite{Guangyuan2017, Piao2017},
\\ \cite{Bhattacharya:2014:IUI:2645710.2645765} \end{tabular}    \\ \hline

& 	& \checkmark		&	\begin{tabular}[c]{@{}l@{}} \cite{Spasojevic:2014:LLS:2623330.2623350} \end{tabular}    \\ \hline

\checkmark & \checkmark	& 		&	\begin{tabular}[c]{@{}l@{}} \cite{Chen2010}, \cite{Budak2014}, 
\\ \cite{Karatay2015a}, \cite{Gao:2011:ITU:2052138.2052335}, 
\\ \cite{Kanta2012}  \end{tabular}    \\ \hline

\checkmark & 	& \checkmark		&	\begin{tabular}[c]{@{}l@{}} \cite{Piao2016d}, \cite{GarciaEsparza:2013:CCT:2449396.2449402}, 
 \\ \cite{Abel2011d, Abel2012a, Abel2011g, Abel2011e, Abel:2013:TUM:2540128.2540558}, \cite{Orlandi2012},
 \\ \cite{Kapanipathi2011}, \cite{Hannon2012}   \end{tabular}    \\ \hline

\end{tabular}
\end{table}
\end{landscape}
}

Although there have been many information sources used for inferring user interests, the comparison of different data sources for inferring user interest profiles has been less explored. Some approaches have utilized different aspects of information of followees such as \emph{topical followees, biographies}, or \emph{list memberships} \citep[e.g.,][]{Besel:2016:ISI:2851613.2851819, Besel:2016:QSI:3015297.3015298, Hannon2012, Bhattacharya:2014:IUI:2645710.2645765, Guangyuan2017}. However, it has not been clearly shown in these studies if these approaches perform better than exploiting users' posts. The usefulness of user interest profiles built from various information sources might be different depending on different applications. For instance, \cite{Chen2010} showed that user interest profiles based on the user's own streams perform better than profiles based on followee streams in the context of URL recommendations on Twitter. However, those profiles based on followee streams might be more useful for recommending followees. 

In addition, combining different information sources have shown its efficiency in a few studies \citep[e.g.,][]{Abel2012a, Piao2017}. However, how to combine different information sources for inferring user interests, and whether there is a synergistic effect on application performance by the combination might require more study. For instance, user interests extracted from different data sources can be either aggregated into a single user interest profile \citep[e.g.,][]{Orlandi2012, Abel2012a} or remain as separate profiles \citep[][]{Piao2017} to measure the preference score of a candidate item for recommendations. Also, combining different data sources has mainly been studied for aggregating user interests from multiple OSNs. Instead, combining different data sources inside the target platform might be useful for inferring user interests as well, e.g., combining extracted user interests from different information sources of followees and users.




\section{Representation of User Interest Profiles}

\subsection{Overview}

In this section, we provide an overview of how user interest profiles have been represented in the different approaches. Here we first provide an overview of user representations for personalized information access that was introduced in \cite{Gauch2007}, and \emph{multi-faceted profiles} which have been proposed in several studies in the literature. We then carry out the review based on three different types of representations in the context of inferring user interest profiles in OSNs in the literature, which include (1) \emph{keyword profiles}, (2) \emph{concept profiles}, and (3) \emph{multi-faceted profiles}.

In \cite{Gauch2007}, the authors defined three types of user representations for personalized information access: 

\begin{itemize}
	\item keyword profiles;
	\item concept profiles;
	\item semantic network profiles.
\end{itemize}

\textbf{Keyword profiles.} In this representation of user interest profiles, each \emph{keyword} or a \emph{group of keywords} can be used for representing a topic of interest. This approach was predominant in every adaptive information retrieval and filtering system and is still popular in these areas \citep{Brusilovsky2007}. When using each keyword for representing user interests, the importance of each word  with respect to users can be measured using a defined weighting scheme such as TF$\cdot$IDF (Term Frequency $\cdot$ Inverse Document Frequency) from information retrieval \citep{Salton1986}. In the case of using groups of keywords for representing user interests, the user interest profiles can be represented as a probability distribution over some topics, and each topic is represented as a probability distribution over a number of words. The topics can be distilled using topic modeling approaches such as Latent Dirichlet Allocation (LDA) \citep{Blei2003}, which is an unsupervised machine learning method to learn topics from a large set of documents. 

\textbf{Concept profiles.} Concept-based user profiles are represented as conceptual nodes (concepts) and their relationships, and the concepts usually come from a pre-existing knowledge base \citep{Gauch2007}. They can be useful for dealing with the problems that keyword profiles have. For example, WordNet \citep{Miller1995} groups related words together in concepts called \emph{synsets}, which has been proved useful for dealing with \emph{polysemy} in other domains. For example, \cite{Stefani} used WordNet synsets for representing user interests in order to provide personalized website access instead of using keywords as they are often not enough for describing someone's interests. Another type of concept is \emph{entities with URIs} (Uniform Resource Identifiers). For instance, this involves using \texttt{dbr\footnote{The prefix \texttt{dbr} denotes http://dbpedia.org/resource/.}:Apple\_Inc.} to denote the company \texttt{Apple}, which is disambiguated based on the context of the word \emph{apple} in a text such as tweet and linked to knowledge bases such as Wikipedia or DBpedia \citep{Auer2007}. DBpedia is the semantic representation of Wikipedia and it has become one of the most important and interlinked datasets on the Web of Data, which indicates a new generation of technologies responsible for the evolution of the current Web from a Web of interlinked documents to a Web of interlinked data \citep{Heath2011}. To facilitate reading, we use DBpedia concepts to denote concepts from Wikipedia or DBpedia.


\textbf{Semantic network profiles.} This type of profile aims to address the polysemy problem of keyword-based profiles by using a weighted semantic network in which each node represents a specific word or a set of related words. This type of profile is similar to concept profiles in the sense of the representation of conceptual nodes and the relationships between them, despite the fact that the concepts in semantic network profiles are learned (modeled)  as part of user profiles by collecting positive/negative feedback from users \citep{Gauch2007}. As most  previous works have focused on implicitly constructing user interest profiles in microblogging services, this type of profile has not been used in the domain of user modeling in microblogging services.

\textbf{Multi-faceted profiles.} Based on these representation strategies, user interest profiles can include different aspects of user interests such as interests inferred from their tweets, profiles or list memeberships. These different aspects of user interests can be combined to construct a single user interest profile or maintained separately as several user interest profiles for a target user. Although it is common to use a single representation with respect to a user interest profile, the \emph{polyrepresentation theory} \citep{Ingwersen1994} based on a cognitive approach indicates that the overlaps between a variety of aspects or contexts with respect to a user within the information retrieval process can decrease the uncertainty and improve the performance of information retrieval. Based on this theory, \cite{White:2009:PUI:1571941.1572005} studied polyrepresentation of user interests in the context of a search engine. The authors combined five different aspects/contexts of a user for inferring user interests, and showed that polyrepresentation is viable for user interest modeling.

\subsection{Review}

\subsubsection{Keyword profiles}

Similar to other adaptive information retrieval and filtering systems, representing user interests using \emph{keywords} or \emph{groups of keywords} is popular in OSNs as well. For instance, \cite{Chen2010} and \cite{Bhattacharya:2014:IUI:2645710.2645765} represented user interest profiles by using vectors of weighted keywords from the tweets and the descriptions of list memberships of users, respectively. Despite the huge volume of information from UGC, extracting keywords from microblogs for inferring user interest profiles is challenging due to the nature of short and noisy messages \citep{Liao2012}.

As an alternative approach, another special type of keyword such as \emph{tags} and \emph{hashtags}\footnote{\url{https://en.wikipedia.org/wiki/Hashtag}} has been used for inferring user interest profiles. In contrast to the words mined from the short texts of microblogs, keywords from tags/hashtags might be more informative and categorical in nature. \cite{Abel2011g, Abel2011d} investigated hashtag-based user interest profiles by extracting hashtags from the tweets of users, and \cite{Hannon2012} leveraged keywords from the tags of users' list memberships for representing their interest profiles. 

Topics distilled from topic modeling approaches such as LDA are also popular for representing user interest profiles. A topic has associated words with their probabilities with respect to the topic. For example, an information technology-related topic can have some top associated words such as ``google, twitter, apple, web''. \cite{Weng:2010:TFT:1718487.1718520} used LDA to distill 50 topics and represented each user as a probability distribution over these topics. In \cite{Abel2011e, Abel2011g, Abel:2013:TUM:2540128.2540558}, the authors also used topics for representing user interests where those topics were extracted by ready-to-use NLP (Natural Language Processing) APIs such as OpenCalais\footnote{\url{http://www.opencalais.com/}}. 
\\\\
\textbf{Pros and cons.} Keyword profiles are the simplest to build, and do not rely on external knowledge from a knowledge base. One of the drawbacks of the keyword-based user profiles is \emph{polysemy}, i.e., a word may have multiple meanings which cannot be distinguished by using keyword-based representation. In addition, these keyword-based approaches lack semantic information and cannot capture relationships among these words, and the assumption of topic modeling approaches that a document has rich information is not the case for microblogs \citep{Zarrinkalam2015}. \cite{Spasojevic:2014:LLS:2623330.2623350} further pointed out that topic modeling approaches cannot provide a scalable solution for inferring topics for millions of users which include a great number of passive users. 

\subsubsection{Concept profiles}

To address some problems of keyword-based approaches, researchers have proposed leveraging \emph{concepts} from KBs such as DBpedia for representing user interests. One of the advantages of leveraging KBs is that we can exploit the background knowledge of these concepts to infer user interests which might not be captured if using keyword-based approaches. For instance, a big fan of the \texttt{Apple} company would be interested in any brand-new products from \texttt{Apple} even the names of these products have never been mentioned in the user's primitive interests \citep{Lu2012}. Concepts from various types of KBs have been leveraged for different purposes of user modeling, such as the ones from simple concept taxonomies with respect to news \citep{Kang2016}, domain-specific KBs such as STW\footnote{\url{http://zbw.eu/stw}}, ACM CCS, and Medical Subject Headings\footnote{\url{https://www.nlm.nih.gov/mesh/}} (MeSH) \citep{Nishioka:2016:PVT:2910896.2910898, Nishioka:2015:ITU:2809563.2809601, Bolting2015}, and cross-domain KBs such as DBpedia \citep{Lu2012, piao2016exploring, Guangyuan2017, Piao2017, Faralli2015, Piao2016b, Piao2016d, Abel2011g, Abel2011d, Abel2011e}. In the following, we discuss some details of the representation strategy using DBpedia concepts which have been the most widely used for representing user interest profiles.


\textbf{Entity-based profiles.} This approach extracts entities from information sources such as a user's tweets, and uses these entities to represent user interest profiles. Take the following real-word tweet as an example \citep{Michelson2010}:
\\\\
``\emph{\#Arsenal winger Walcott: Becks is my England inspiration: http://tinyurl.com/37zyjsc}'',
\\\\
there are four entities such as \texttt{dbr:Arsenal\_F.C.}, and \texttt{dbr:Theo\_Walcott} within the tweet, which can be used for constructing entity-based user interest profiles. However, this approach is difficult to infer more specific interests which might need to be represented by combining multiple related entities or interests that cannot be found in a knowledge base. To address this issue, some studies have proposed representing each topic of interest as a \emph{conjunction of multiple entities}, which are correlated on Twitter in a certain timespan \citep{Zarrinkalam2015a, Zarrinkalam2016}. These sets of entities for representing a topic of interest can be learned via unsupervised approaches in a similar manner to learning topics with topic modeling approaches for keyword-based profiles.

\begin{figure*}[!h]
\includegraphics[angle=90, width=\textwidth]{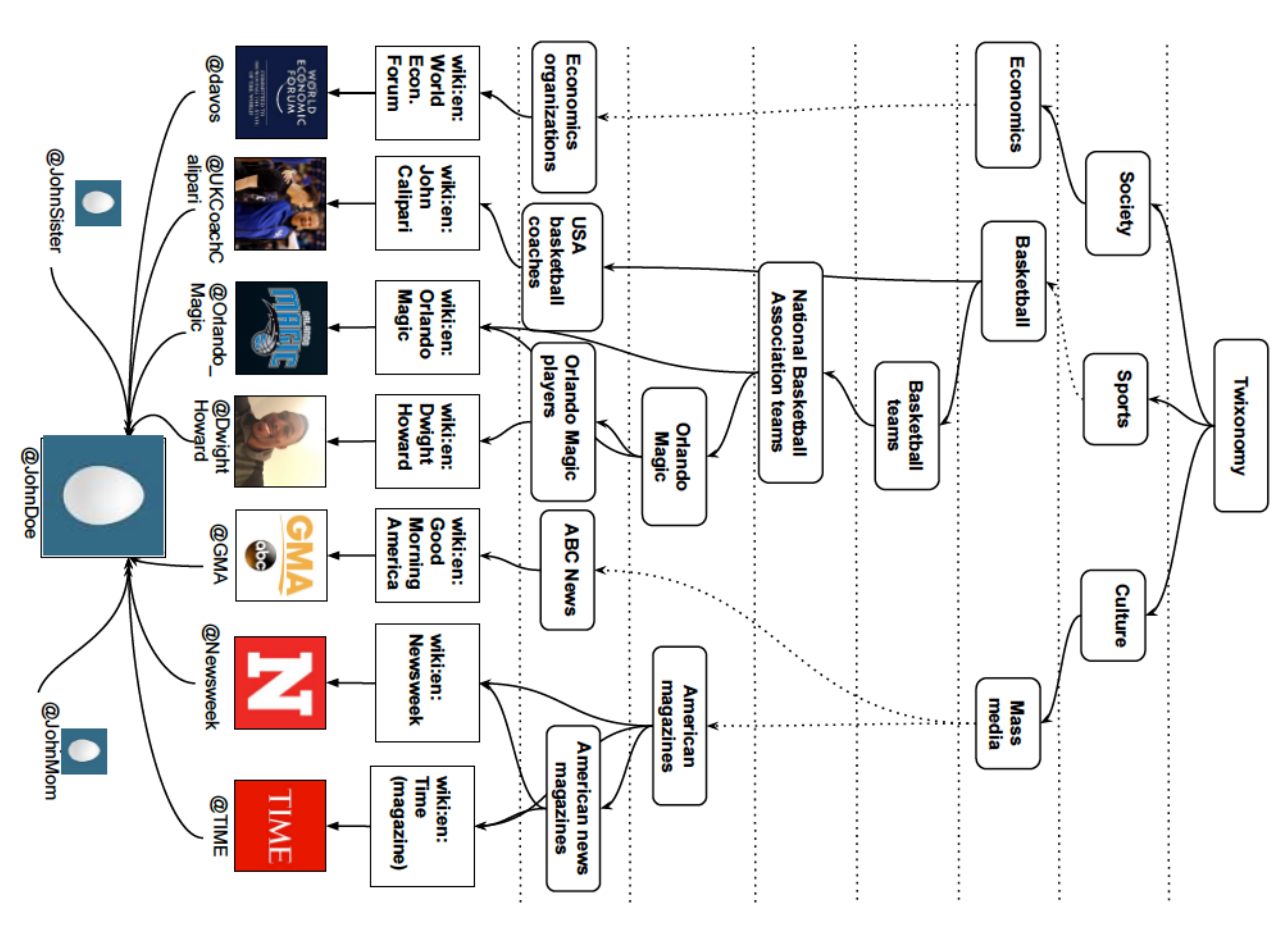}
\caption{An example of Twixonomy for a single user \citep{Faralli2017}.}
\label{twixonomy}       
\end{figure*}

\textbf{Category-based profiles.} An alternative approach is using DBpedia \emph{categories}, which represents more general user interests compared to using DBpedia \emph{entities}. Returning to the example in the previous paragraph, the categories of the mentioned entities in that tweet such as \texttt{dbr:Category:English\_Football\_League} can be used for representing the topic of interests instead of those entities. One can also choose the level or depth of categories in a KB for representing user interest profiles or use all categories related to primitive interests. The top-level DBpedia categories can refer to general ones such as \texttt{dbr:Category:Sports} and \texttt{dbr:Category:Health} compared to the categories in a lower level such as \texttt{dbr:Category:English\_Football\_League}. For example, \cite{Michelson2010} and \cite{Nechaev} used top-level categories to represent user interest profiles while other studies \citep[][etc.]{Faralli2017, Kapanipathi2014, Flati2014} used hierarchical categories to represent user interest profiles. Figure \ref{twixonomy} shows an example of category-based representation of user interests based on extracted entities from followees' account names, which is called \emph{Twixonomy} \citep{Faralli2017}. 

\textbf{Hybrid representations.} Each aforementioned representation has its strengths and weaknesses. In terms of entity- or category-based representations, extracting entities with URIs is a fundamental step for constructing either \emph{entity-} or \emph{category-based} user interest profiles. However, the task of extracting entities is non-trivial \citep{Kapanipathi2014} due to the noisy, informal language of microblogs \citep{Ritter2011}. In addition, knowledge bases might be out-of-date for emerging concepts on microblogging services, and therefore cannot capture these concepts during the entity extraction process. To overcome the drawbacks of using a single interest format, \emph{hybrid representations} based on various interest formats have been explored as well. Instead of using only entities or categories for representing user interests, hybrid approaches combine different interest formats for constructing user profiles \citep{piao2016exploring, Guangyuan2017, Piao2017, Faralli2015, Piao2016d, Nishioka:2016:PVT:2910896.2910898, OBanion2012}. For example, \cite{OBanion2012} used categories as well as entities to represent user interest profiles. \cite{Piao2016d, Piao2016b} proposed a hybrid approach using both DBpedia entities and WordNet synsets for representing user interests in order to capture user interests that might be missed due to the problem with entity recognition in microblogs.
\\\\
\textbf{Pros and cons.} On the one hand, concept-based approaches present the semantics between concepts and can leverage background knowledge about concepts for propagating user interest profiles. On the other hand, these approaches rely on pre-existing or pre-constructed KBs which might be not always available in or lack of coverage with respect to some domains.

\subsubsection{Multi-faceted profiles}

Multi-faceted profiles model multiple aspects for a target user based on different information sources or using different representation strategies in order to derive a comprehensive view of that user. The assumption here is that different aspects of users may complement each other and improve the inferred user interest profiles. 

\begin{figure*}[!b]
\includegraphics[width=\textwidth]{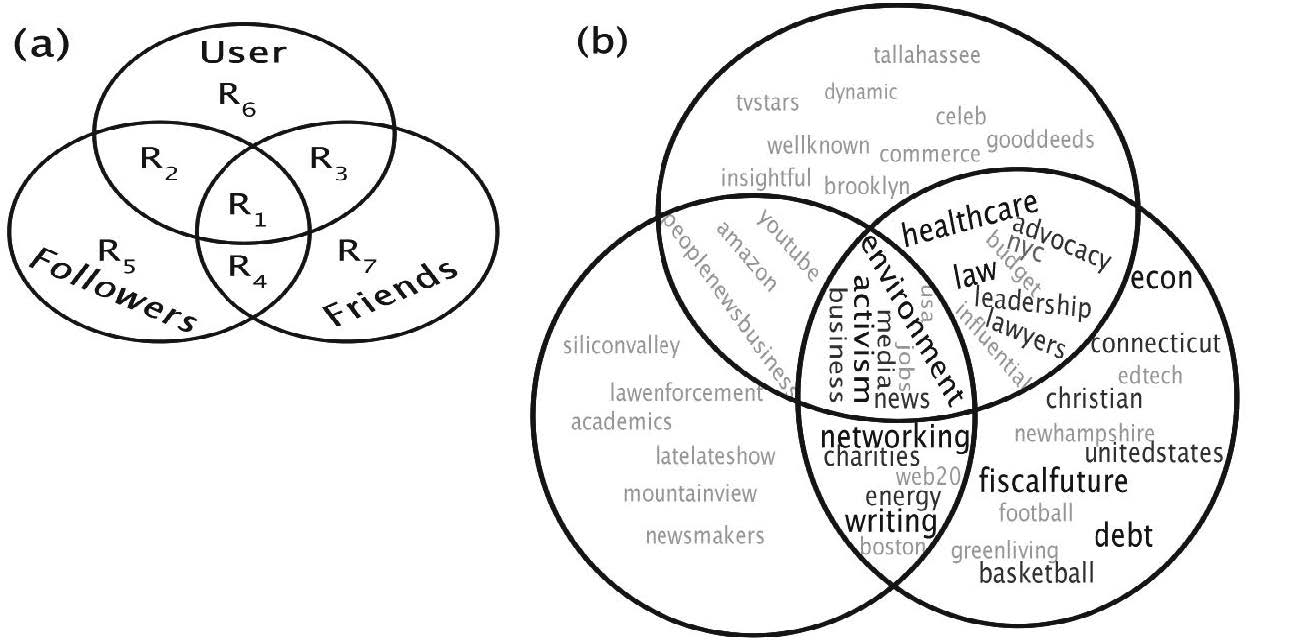}
\caption{(a) Intensional and extensional profile regions. (b) Barack Obama's profile showing the tags associated with Obama and his followees (friends in the figure) and followers \citep{Hannon2012}.}
\label{multi}       
\end{figure*}

\cite{Hannon2012} proposed a multi-faceted user profile which includes user interests from target users, their followees, and followers. Figure \ref{multi} shows an example from \cite{Hannon2012} for representing user interests, where user interests are represented based on the tags of list memberships of users, followees, or followers provided by a third-party service. The figure shows that user interests inferred from different aspects can complement each other and lead to a better understanding of a target user. However, they did not evaluate the effectiveness of multi-faceted profiles in the context of personalized recommendations and left it as a future work. 

The authors in \cite{Lu2012} and \cite{Chen2010} both constructed two keyword-based user interest profiles for each user. In \cite{Chen2010}, two keyword-based user interest profiles were built based on the tweets of users and those of their followees for recommending URLs on Twitter. The results in \cite{Chen2010} showed that using user interest profiles based on the tweets of users performs better than using those based on the tweets of their followees. \cite{Lu2012} proposed using DBpedia entities and the affinity of other users to construct two user interest profiles for recommending tweets on Twitter. For a given user, the first user profile was represented as a vector of DBpedia entities, which were extracted from the user's tweets. Both of these studies did not investigate the synergistic effect of combining these two aspects compared to considering a single aspect of users. More recently, \cite{Piao2017} showed that leveraging concept-based profiles from the biographies and list memberships of followees can complement each other and improve the URL recommendation performance on Twitter.
\\\\
\textbf{Pros and cons.} Multi-faceted profiles provide a comprehensive view of a user with respect to his/her interests and can improve recommendation performance. On the other hand, multiple information sources have to be explored for constructing multi-faceted profiles.


\subsection{Summary and discussion}

In this section, we reviewed various ways of representing user interests such as using \emph{keywords}, various types of \emph{concepts}, and some multi-faceted approaches. Table \ref{representation} shows a summary of different representations of user interests adopted by previous studies.

Those different representations of user interests might work differently depending on the application where these user profiles are used. For example, we usually have to construct item profiles in the same way as constructing user interest profiles in order to measure the similarity between them for providing recommendations. The entity-based representation strategies for user interests might be appropriate for recommending items with long content, e.g., news or URL recommendations as the content of them is usually long. In contrast, these representation strategies might not work well for recommending items with short descriptions such as tweets due to the difficulty of extracting entities from them. 
For example, the low recall of entities on Twitter has been reported in both \cite{Kapanipathi2014} and \cite{Piao2016d} using several state-of-the-art NLP APIs. In a recent study \citep{Manrique:2017:SDA:3106426.3109440}, the authors also showed that 30\% of the titles of a research article cannot extract any entity at all. Some hybrid approaches such as combining word- and concept-based representations might be useful in this case.

In addition, different facets should be considered carefully for constructing multi-faceted profiles in the context of item recommendations. Each facet of multi-faceted profiles can have different importance for the recommended items, and leveraging completely unrelated facets might introduce noise to the constructed profiles. For example, \cite{Piao2017} showed that different weights are required for different facets in order to achieve the best performance in URL recommendations on Twitter. \cite{Abel2013} showed that it is helpful to have sufficient overlap between different facets of multi-faceted profiles for tag recommendations in a cold start.

\renewcommand{\arraystretch}{1.5}
\afterpage{
\begin{landscape}
\begin{table}[!h]
\centering
\caption{Representation of user interest profiles.}
\label{representation}
\begin{tabular}{|c|c|c|l|}
\hline
\textbf{\begin{tabular}[c]{@{}l@{}}Keyword \\Profiles\end{tabular}} & \textbf{\begin{tabular}[c]{@{}l@{}}Concept \\Profiles\end{tabular}} & \textbf{\begin{tabular}[c]{@{}l@{}}Multi-faceted \end{tabular}}  & \textbf{Examples} \\ \hline

\checkmark &  & \checkmark & \cite{Chen2010}, \cite{Hannon2012}     \\ \hline

 & \checkmark & \checkmark & \cite{Lu2012}, \cite{Piao2017}, \cite{Spasojevic:2014:LLS:2623330.2623350}    \\ \hline
 
 \checkmark  & \checkmark & &  \begin{tabular}[c]{@{}l@{}} \cite{Abel2011g, Abel2011e, Abel2011d, Abel:2013:TUM:2540128.2540558}, \cite{Kanta2012} \end{tabular} \\ \hline

\checkmark  & &  & \begin{tabular}[c]{@{}l@{}}  \cite{Weng:2010:TFT:1718487.1718520}, \cite{Xu2011}, 
\\ \cite{Sang:2015:PFT:2806416.2806470}, \cite{Bhattacharya:2014:IUI:2645710.2645765}, 
\\ \cite{Vu:2013:IMU:2505515.2507883}, \cite{Phelan:2009:UTR:1639714.1639794}  \end{tabular}    \\ \hline
 
  & \checkmark & &  \begin{tabular}[c]{@{}l@{}} \cite{Kapanipathi2014}, \cite{Besel:2016:ISI:2851613.2851819, Besel:2016:QSI:3015297.3015298}, \cite{Faralli2017}, \\ \cite{Siehndel:2012:TUP:2887379.2887395}, \cite{Michelson2010}, 
  \\ \cite{piao2016exploring, Guangyuan2017, Piao2016b, Piao2016d}, \cite{Karatay2015a},
  \\ \cite{Kang2016}, \cite{Abel2012a}, \cite{Narducci2013},  
  \\ \cite{Orlandi2012}, \cite{Kapanipathi2011}, \cite{Jipmo2017},
  \\ \cite{Zarrinkalam2015a}, \cite{Zarrinkalam2016}, 
  \\ \cite{Bhargava:2015:UMU:2678025.2701365}, \cite{GarciaEsparza:2013:CCT:2449396.2449402}, 
  \\ \cite{Nishioka:2016:PVT:2910896.2910898,Nishioka:2015:ITU:2809563.2809601}, \cite{Jiang2015},  
  \\ \cite{Bolting2015}, \cite{Gao:2011:ITU:2052138.2052335}, \cite{Nechaev},
  \\ \cite{Budak2014}, \cite{Penas2013}, \cite{AnilKumarTrikhaFattaneZarrinkalam},
  \\ \cite{OBanion2012}, \cite{Ahn:2012:IUI:2457524.2457681}, \cite{Lim:2013:ICT:2491055.2491078} \end{tabular}    \\ \hline

\end{tabular}
\end{table}
\end{landscape}
}

It is also worth noting that the structure of user interest profiles can be different even with the same user interest format. Take a category-based user interest profile as an example, it can be a \emph{vector}, \emph{taxonomy} or \emph{graph} by retaining the hierarchical or general relationships among categories. Also, the final profile extracted from the same structure can be different. For instance, both user interest profiles proposed in \cite{Faralli2017} (see Figure \ref{twixonomy}) and \cite{Kapanipathi2014} were represented as a \emph{taxonomy} at first, but were used differently for the final representation of user interests. In \cite{Faralli2017}, entities or categories in different levels were used separately as an interest vector for representing a user, e.g., using categories that were two hops away from the user's primitive interests as the final interest profile. However, using a specific abstraction level of the category taxonomy for all users does not consider that different users might have different depths or expertise levels in terms of a topic of interests. In contrast, \cite{Kapanipathi2014} sorted all categories in the taxonomy of a user based on their weights for representing the user's interest profile. The different usages of the category taxonomy indicate some opportunities and challenges. On the one hand, the taxonomy structure of user interests is flexible enough to extract different abstraction levels of user interests or an overview of them. On the other hand, it has not been investigated which type of user interest profile obtained from the taxonomy structure is better.  



\section{Construction and Enhancement of User Interest Profiles}
\subsection{Overview}
So far we have focused our discussion on collecting data from various sources for inferring user interests, and different representations for interest profiles. In this section, we provide details on how user interest profiles of a certain representation can be constructed based on the collected data. The overview of the construction and enhancement of user interest profiles is carried out based on three criteria:

\begin{itemize}
	\item profile construction with weighting schemes;
	\item profile enhancement;
	\item temporal dynamics of user interests.
\end{itemize}

Based on a defined representation of user interest profiles, a profile constructor aims to determine the weights of user interest formats such as words or concepts in user profiles with a certain \emph{weighting scheme}. The weights of interest formats denote the importance of these interests with respect to a user. In Section \ref{Profile Construction}, we review different weighting schemes based on various information sources such as users' posts or their followees, etc.

Primitive interest profiles, e.g., entity-based user profiles, can be further enhanced by using background knowledge from knowledge bases. For instance, this can be achieved by inferring category-based user interest profiles on top of the extracted entities from the data collected. Section \ref{Profile Enhancement} describes the approaches leveraging knowledge bases for enhancing primitive interest profiles.

User interests can change over time in OSNs. For instance, a user interest profile built during the last two weeks might be totally different from one built from two years ago. In Section \ref{Temporal Dynamics of User Interests}, we look at whether or not the temporal dynamics of user interests have been considered when constructing user interest profiles, and if yes, how they have been incorporated during the construction process.

\subsection{Review}

\subsubsection{Profile Construction with Weighting Schemes}
\label{Profile Construction}

The output of a profile constructor is a primitive user interest profile represented by weighted interests based on a predefined representation. A \emph{weighting scheme} is a function or process to determine the weights of user interests. 

\textbf{Heuristic approaches.} A common and simple weighting scheme is using the frequency of an interest $i$ (e.g., a keyword or an entity) to denote the importance of $i$ with respect to a user \emph{u}, which can be formulated as below when the data source is \emph{u}'s posts: 

\begin{equation}
	TF_u(w_i) = frequency\mbox{ }of\mbox{ }i\mbox{ }in\mbox{ }u's\mbox{ }posts.
\end{equation}
\\
Despite its simplicity, this approach has been widely used in the literature, particularly in entity-based user interest representations \citep{Kapanipathi2014, Abel2011e, Tao2012}. Interests represented as concepts such as entities extracted from tweets might come with their confidence scores, and these scores can be incorporated into a weighting scheme. For instance, \cite{Jiang2015} used TF with the confidence scores of extracted entities from tweets as their weighting scheme. 

One problem with TF is that common words or entities which appear frequently in many users' interest profiles and may not be important as user interests. TF$\cdot$IDF is another common weighting scheme to cope with this problem. The IDF score of $i$ with respect to a user \emph{u} based on \emph{u}'s tweets can be measured as below \citep{Chen2010}:

\begin{equation}
	IDF_u(i) = log\left[\frac{\#\mbox{ }all\mbox{ }users}{\#\mbox{ }users\mbox{ }using\mbox{ }i\mbox{ }at\mbox{ }least\mbox{ }once}\right].
\end{equation}
\\
Instead of using users for measuring the IDF score of an interest, IDF has been applied in other ways as well. For example, \cite{Nishioka:2016:PVT:2910896.2910898} applied IDF with randomly retrieved tweets from the streaming API of Twitter, and \cite{Gao:2011:ITU:2052138.2052335} applied IDF to value the specificity of an interest within a given period of time. It is worth noting that the IDF weighting can also be applied after the \emph{profile enhancement} process \citep[e.g.,][]{Piao2016d, Nishioka:2016:PVT:2910896.2910898}. 

More sophisticated approaches can be applied for weighting user interests. In \cite{Vu:2013:IMU:2505515.2507883}, the authors compared different weighting schemes such as TF$\cdot$IDF, TextRank \citep{mihalcea-tarau:2004:EMNLP}, and TI-TextRank which was proposed by the authors by combining TF$\cdot$IDF and TextRank. Based on a user study, the authors showed that TI-TextRank performs best for ranking keywords from the tweets of users. 


In the context of OSNs, specific approaches have to be devised for constructing user interest profiles by exploiting their social networks such as followees on Twitter \citep{Chen2010, Lu2012}. To this end, several methods have been proposed. For example, \cite{Chen2010} first retrieved a set of \emph{high-interest words} for followees as follows in order to build a user profile based on followees' tweets: First, keyword-based user interest profiles were created using the TF$\cdot$IDF weighting scheme based on the tweets of followees, which are called \emph{self-profiles}. Next, for each \emph{self-profile} for followees of \emph{u}, they picked all words that have been mentioned at least once, and selected the top 20\% of words based on their occurrences. In addition, the words that are not in other followees' profiles were removed. Subsequently, the weight of each word in the set of \emph{high-interest words} was measured as below:

\begin{equation}
\begin{split}
	FTF_u(i) = \mbox{ }& \#\mbox{ }u's\mbox{ } followees\mbox{ } who\mbox{ } have\mbox{ } i\mbox{ }\\
			      \mbox{ }&  as\mbox{ } one\mbox{ } of\mbox{ } their\mbox{ } high-interest\mbox{ } words.
\end{split}
\end{equation}
\\
Similar approaches of $FTF_u(i)$ were adopted in \cite{Piao2017} and \cite{Bhattacharya:2014:IUI:2645710.2645765} but by exploring the list memberships of followees instead of their tweets for extracting user interests. 

An alternative approach for aggregating the weights of interests in the followees' profiles is normalizing each followee's profiles and then aggregating those normalized weights for building user interest profiles \citep{Piao2017, Spasojevic:2014:LLS:2623330.2623350}. In \cite{Piao2017}, the authors showed that this simple alternative approach performs better compared to $FTF_u(i)$ for weighting entities extracted from the list memberships of followees when using inferred user interest profiles for URL recommendations on Twitter. These approaches assume that each followee is equally important when aggregating their interest profiles for building the user interest profile of a target user. However, some followees' profiles can be more important compared to others with respect to the target user. In \cite{Karatay2015a}, the authors incorporated the relative ranking scores of social networks into their weighting scheme to weight the entities of users.

\textbf{Probablistic approaches.} The aforementioned approaches focus on interests such as entities appearing in users' posts, however, not all the entities related to a post explicitly appear in that post. In this regard, some approaches extracted interests such as entities by measuring the similarity between a post and an entity. For instance, \cite{Lu2012} and \cite{Narducci2013} used the Explicit Semantic Analysis (ESA) \citep{Gabrilovich} algorithm, which is designed to compute the similarity between texts, for obtaining the weights of entities for each tweet of a user. Those weights of entities were then aggregated for constructing entity-based primitive interests of users. \cite{Ahn:2012:IUI:2457524.2457681} quantified the degree of an interest, i.e., a Facebook entity, based on two factors: (1) the familiarity with each social neighbor, and (2) the similarity between the topic distributions of a social content and an interest. \emph{Social content} is the combined text of a post and its comments between users, and the topic distributions of it is obtained using LDA.

%
%
 
The weights of user interests have also been learned in unsupervised ways in the literature. For instance, \cite{Weng:2010:TFT:1718487.1718520} treated tweet histories of each user as a big document, and used LDA to learn topic distributions for each user. \cite{AnilKumarTrikhaFattaneZarrinkalam} and \cite{Zarrinkalam2017} also used LDA to infer topic distributions for each user in time intervals where a topic is a set of DBpedia entities. Similarly, user interest profiles were represented as topic vectors where each topic is a set of temporally correlated entities on Twitter in \cite{Zarrinkalam2015a}. To this end, an entity graph based on their temporal correlation as defined by the authors was constructed, and the topics in a time interval were extracted using some existing community detection algorithms such as the \emph{Louvain} method \citep{Rotta:2011:MLS:1963190.1970376}. The Louvain method is a simple and efficient algorithm for community detection, and relies upon a heuristic for optimizing modularity which quantifies the density of the links inside of the communities as compared to the links between communities. Subsequently, each topic $z$ was transformed into a set of weighted entities using the \emph{degree centrality} of an entity in the topic (community). Finally, they obtained the weight of a topic based on the weight of an entity \emph{c} with respect to the topic and the frequency of \emph{c} in \emph{u}'s tweets.
%
%

\cite{Budak2014} proposed a probabilistic generative model to infer user interest profiles which are represented as an interest probability distribution over ODP (Open Directory Project\footnote{\url{https://en.wikipedia.org/wiki/DMOZ}}) categories. In their proposed approach, the authors considered three aspects such as (1) the posts of a target user, (2) the activeness of the user, and (3) the influence of friends. They assumed that time is divided into fixed time steps, and transformed the problem into inferring the probability of a user being interested in each of the interests, given a social network that evolves over time including posts and social network information. 
%
\cite{Sang:2015:PFT:2806416.2806470} also proposed a probabilistic framework for inferring user interest profiles. Differring from \cite{Budak2014}, \cite{Sang:2015:PFT:2806416.2806470} assumed users have long- and short-term interest (topic) distributions. Long-term interests denote stable preferences of users while short-term interests denote user preferences over short-term topics of events in OSNs. However, they did not consider users' social networks. 

In contrast to the aforementioned approaches, which assume all tweets posted by users are related to their interests, \cite{Xu2011} proposed a modified author-topic model \citep{Rosen-Zvi:2004:AMA:1036843.1036902} for distinguishing interest-related and unrelated tweets when learning the topic distributions of users.

\subsubsection{Profile Enhancement}
\label{Profile Enhancement}

One of the advantages of constructing primitive interest profiles using concepts such as entities is that they can be further enhanced by external knowledge to deliver the final interest profiles. The approaches used in the literature for enhancing primitive user interests have mainly leveraged  \emph{hierarchical}, \emph{graph-based}, or \emph{collective} knowledge.

\textbf{Leveraging hierarchical knowledge}. One line of approach for enhancing entity-based primitive interest profiles is apply an adapted \emph{spreading activation} \citep{Collins1975} function on a hierarchical knowledge base. For example, \cite{Kapanipathi2014} proposed representing user interest profiles as Wikipedia categories based on a hierarchical knowledge base, which is a refined Wikipedia category system built by the authors. 
The user interest profiles were then constructed using the hierarchical knowledge base with the following two steps. First, Wikipedia entities in users' tweets were extracted as their primitive interests. Second, these entities were used as activated nodes for applying an adapted spreading activation function on the hierarchical knowledge base in order to infer weighted categories for representing user interest profiles. 

%

The spreading activation function proposed by \cite{Kapanipathi2014} can be applied to any case where a set of entities and a hierarchical knowledge base are available. Therefore, many studies that followed have adopted this function but with different approaches for extracting entities or with different hierarchical knowledge bases \citep{Besel:2016:QSI:3015297.3015298, Besel:2016:ISI:2851613.2851819, Guangyuan2017, Nishioka:2016:PVT:2910896.2910898, Bolting2015}. For instance, \cite{Nishioka:2016:PVT:2910896.2910898} extracted entities and applied the spreading activation function on STW, which is a hierarchical knowledge base from the economics domain. \cite{Bolting2015} investigated several spreading activation functions including the one proposed in \cite{Kapanipathi2014} with the ACM CCS concept taxonomy in the computer science domain. The results showed that using a basic spreading activation function provides the best user interest profiles compared to using other ones in the context of research article recommendations.
%

\cite{Besel:2016:QSI:3015297.3015298,Besel:2016:ISI:2851613.2851819} extracted entities by mapping followees' Twitter accounts to Wikipedia entities, and used WiBi \citep{Flati2014} as their hierarchical knowledge base for applying the spreading activation function proposed in \cite{Kapanipathi2014}. Similarly, \cite{Faralli2015} also mapped followees' Twitter accounts to Wikipedia entities, and used them as users' primitive interests for propagation with WiBi. However, a simpler propagation strategy was adopted in \cite{Faralli2015}. In \cite{Faralli2017}, the authors extended their previous work \citep{Faralli} and proposed a methodology to build \emph{Twixonomy}, which is a Wikipedia category taxonomy. \emph{Twixonomy} is built by using a graph pruning approach based on a variant of Edmonds optimal branching \citep{Edmonds}. The authors showed that the proposed approach can generate a more accurate taxonomy compared to the one proposed in \cite{Kapanipathi2014}. As we mentioned in Section \ref{Using information inside the platform}, one issue with these approaches mapping followees' accounts to Wikipedia entities is that only a limited percentage of followees' accounts can be mapped to corresponding entities. For example, \cite{Faralli2015} and \cite{Guangyuan2017} reported that only 12.7\% and 10\% of followees' accounts can be mapped to Wikipedia entities. In this regard, \cite{Guangyuan2017} considered the use of followees' \emph{biographies} for extracting entities, and applied two different propagation strategies; one is the spreading activation function from \cite{Kapanipathi2014}, and the other is an interest propagation strategy exploring the DBpedia knowledge graph which will be discussed later on \citep{piao2016exploring}. 

Instead of using refined hierarchical knowledge from Wikipedia, some studies have explored other types of hierarchical knowledge bases as well. \cite{Kang2016} proposed mapping news categories to tweets for constructing user interest profiles. The authors leveraged news categories from two popular news portals in South Korea (Naver News\footnote{\url{http://news.naver.com/}} and Nate News\footnote{\url{http://news.nate.com//}}) to build their category taxonomy. This taxonomy consists of 8 main categories and 58 sub-categories, and each category consists of all news articles in the two news corpuses. To assign categories to a tweet, each tweet and news category are represented as a term vector where the weights of terms are calculated using TF$\cdot$IDF first. As there might be a semantic gap between terms in social media and news portals, the authors leveraged Wikipedia to transform the term vectors of tweets and news categories into a same vector space. The top two news categories to each tweet based on the cosine similarity between their vectors, and these news categories of a user's tweets are then aggregated to construct the final user interest profiles. 

\cite{Jiang2015} leveraged external knowledge sources such as DBpedia, Freebase \citep{Bollacker2008}, and Yago \citep{Suchanek2007a} for constructing a topic hierarchy tree, which is a hierarchical knowledge base consists of over 1,000 topics distributed in 5 levels. However, the details for obtaining the topic hierarchy tree were not discussed in their study. The topic hierarchy tree used in Klout service is also bootstrapped using Freebase and Wikipedia, which consists of 3 levels with 15, around 700, and around 9,000 concepts in each level, respectively \citep{Spasojevic:2014:LLS:2623330.2623350}. In \cite{Bhargava:2015:UMU:2678025.2701365}, the authors manually built a category taxonomy based on Facebook Page categories and the Yelp\footnote{\url{https://www.yelp.com/}} category list. The category taxonomy in \cite{Bhargava:2015:UMU:2678025.2701365} consists of three levels with 8, 58, and 137 categories in each level, respectively. The authors used features such as entities, hashtags, and document categories which can be extracted from Facebook \emph{likes} and UGC as users' primitive interests, and then measured the confidence of each concept in the category taxonomy based on these features using the Semantic Textual Similarity system \citep{Han2013}.


\textbf{Leveraging graph-based knowledge}. Instead of leveraging hierarchical knowledge, many studies have leveraged graph-based knowledge for enhancing user profiles. For example, \cite{Michelson2010} exploited Wikipedia categories directly for propagating a user's primitive interests. The authors summed the scores of a category which appeared in multiple depths in the category graph. Differing from exploring the categories of a specified depth \citep{Michelson2010}, \cite{Siehndel:2012:TUP:2887379.2887395} represented user interest profiles using 23 top-level categories of the root node \texttt{Category:Main\_Topic\_Classifications} in Wikipedia. The Wikipedia entities in users' tweets were extracted as their \emph{primitive interests}, and these entities were then propagated up to the 23 top-level categories with a discounting strategy for the propagation. 

\begin{figure}[!t]
  \centering
  \subfloat[WiBi taxonomy]{\includegraphics[width=0.7\textwidth]{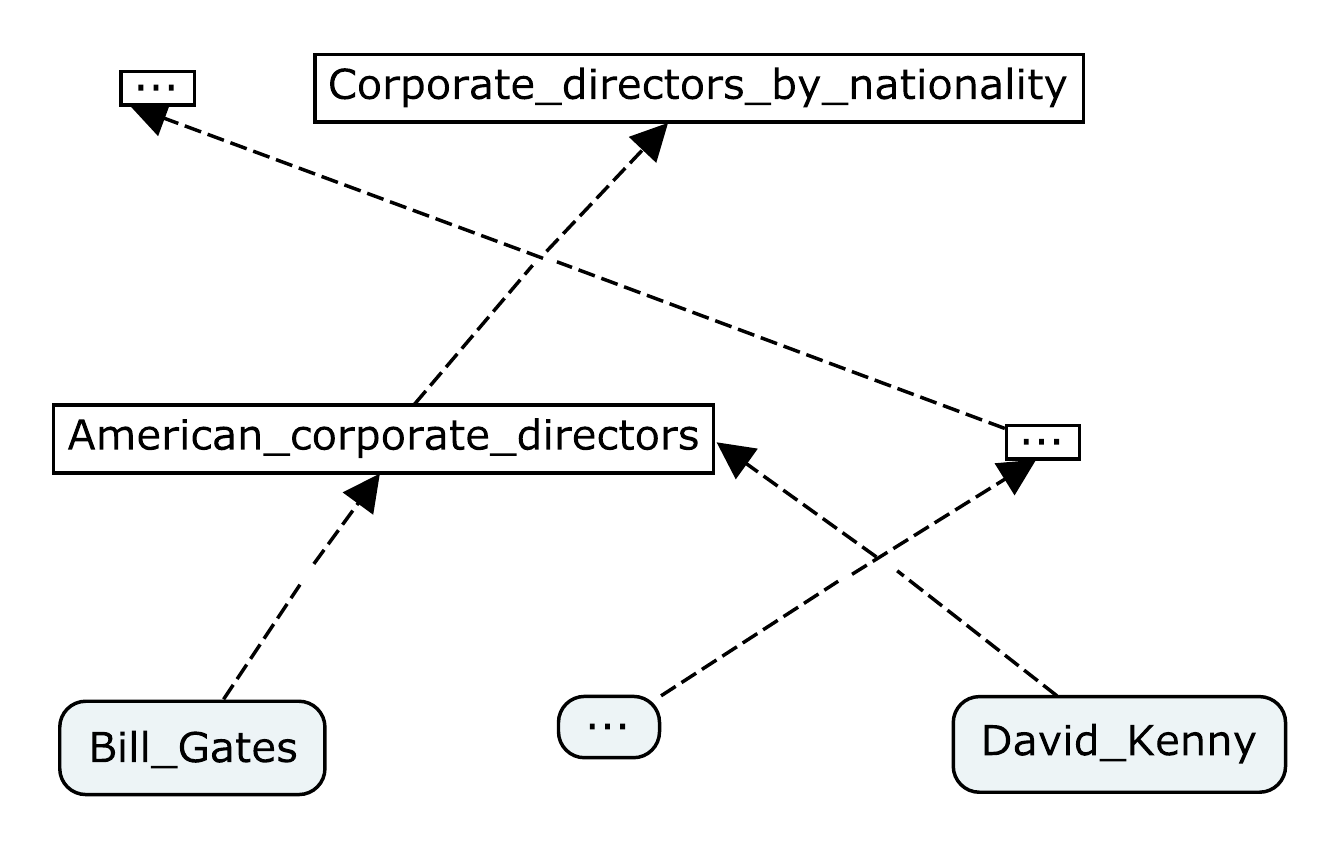}\label{fig:wibi}}
  \hfill
  \subfloat[DBpedia graph]{\includegraphics[width=0.7\textwidth]{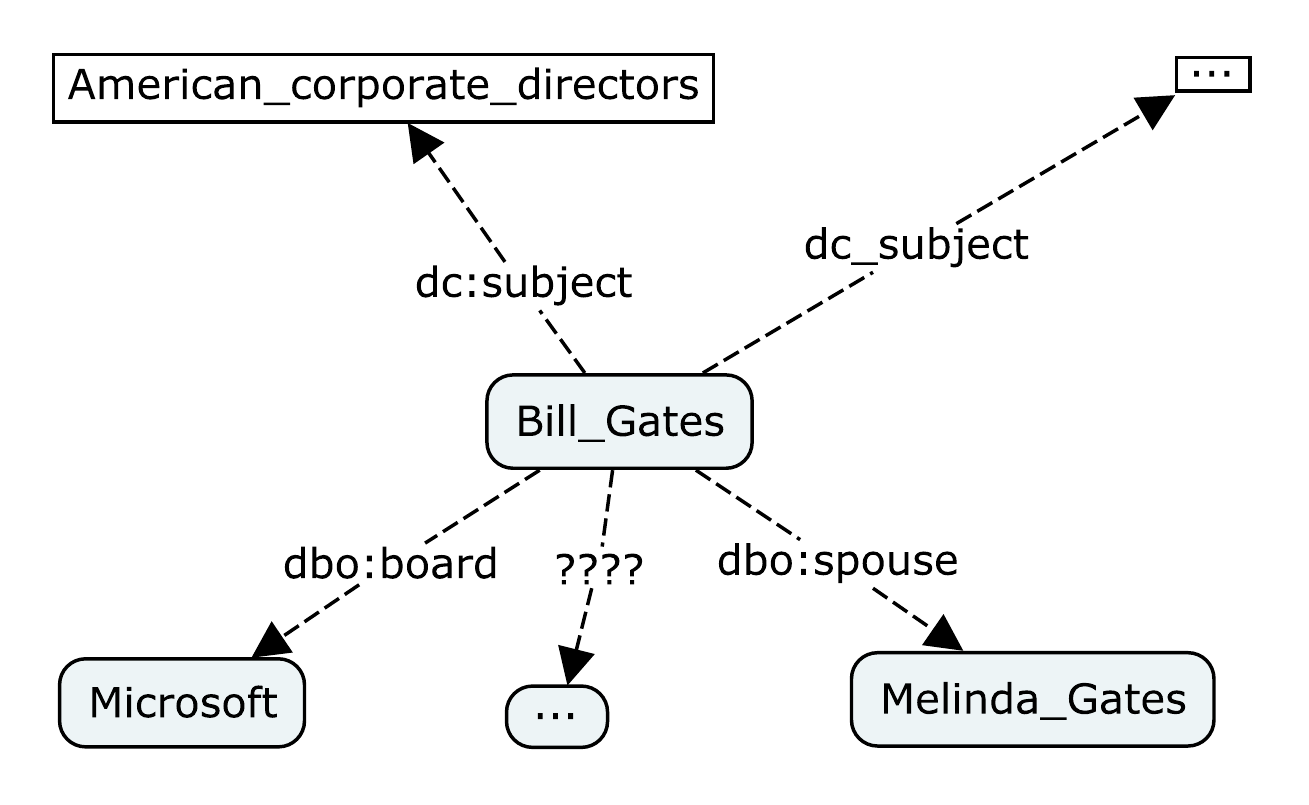}\label{fig:dbpedia}}
  \caption{Examples of WiBi taxonomy and DBpedia graph \citep{Guangyuan2017}.}
  \label{fig:wibidbpedia}
\end{figure}

With the advent of large, cross-domain Knowledge Graphs (KGs) such as DBpedia, different approaches leveraging background knowledge from KGs have been investigated. A knowledge graph is a knowledge base which consists of an ontology and instances of the classes in the ontology \citep{Farber}. The difference between a hierarchical category taxonomy such as WiBi and a knowledge graph such as DBpedia is displayed in Figure \ref{fig:wibidbpedia} \citep{Guangyuan2017}. As we can see from the figure, for an entity, DBpedia goes beyond just categories to provide related entities via the entity's properties/edges. Depending on the propagation strategies for those entities in a user's primitive interests, different aspects, e.g., \emph{related entities}, \emph{categories} or \emph{classes} of the entities can be leveraged for the propagation. For example, \cite{Penas2013} enriched categories in users' primitive interests using similar categories defined by the \texttt{categorySameAs} relationship in DBpedia. \cite{Abel2012a} proposed using background knowledge from DBpedia for propagating user interest profiles with respect to POI. The authors considered entities that were two hops away from a user's primitive interests and that were related to places. However, this approach did not consider any discounting strategy for the weights of propagated user interests. In \cite{Orlandi2012}, the authors leveraged DBpedia categories one hop away from of the entities in a user's primitive interests using a discounting strategy for propagating user interests. 

%
%

Although \cite{Orlandi2012} leveraged DBpedia as the knowledge base instead of Wikipedia, they still exploited categories only, which makes no difference between using DBpedia and Wikipedia. To investigate other aspects of DBpedia such as related entities and classes of primitive interests, \cite{piao2016exploring} studied three approaches such as \emph{category-}, \emph{class-}, and \emph{property-based} propagation strategies. This study found that exploiting categories and related entities via different properties of primitive interests provides the best performance compared to using corresponding categories only in the context of URL recommendations on Twitter. 

%
%

An alternative graph for propagating entity-based user interest profiles is the Wikipedia entity graph. Compared to the DBpedia graph, where the edges between two entities are predefined properties in an ontology, the edges in the Wikipedia entity graph denote the mentions of the other entities in a Wikipedia entity (article). \cite{Lu2012} exploited a Wikipedia entity graph to enhance the entity-based primitive interests. Different from exploiting Wikipedia categories, the intuition behind this approach is that if a user is interested in \texttt{IPhone}, the user might be interested in other products from \texttt{Apple}, instead of being interested in other mobile phones in the same category such as \texttt{Smartphones}. To this end, the authors used the ESA algorithm to extract entities from the tweets of users as their primitive interests, and then expanded these entities using a random walk on the Wikipedia entity graph. 

%

In \cite{Jipmo2017}, the authors assumed there are a set of interests $i \in I$, e.g., \texttt{Sports}, \texttt{Politics}, etc., which the user modeling system needs to measure the corresponding weights for each interest. After building a bag of entities based on the ones extracted from a user's tweets, the relevance score of an interest $i$ is measured as below, which can be seen as a spreading activation approach with some constraints:

\begin{equation}
	S_i^u = \sum_{a \in BOE_u} \frac{1}{min\{dist(a, c), c \in BOC_i\}}
\bigskip
\end{equation}
where $BOE_u$ denotes the bag of entities extracted from $u's$ tweets, and $BOC_i$ denotes a set of categories containing the name of $i$ in their titles. For example, for an interest \texttt{sports}, $BOC_i$ consists of categories such as \texttt{Category:Sports by year, Category:Sports in France}, etc. $dist(a, c)$ refers to the length of the shortest directed path from $a$ to $c$ in the Wikipedia graph. 

\textbf{Leveraging collective knowledge}. More recently, some studies proposed leveraging collective knowledge powered by the great amount of interest profiles of all users in a dataset, and enhancing a user profile with other related interests identified as frequent patterns in all profiles using frequent pattern mining (FPM). FPM was designed to find frequent patterns (itemsets or a set of items that appear together in a transaction dataset frequently). In the context of user modeling, previous studies have treated each user interest as an item, each interest profile as a transaction, and all user interest profiles as the transaction dataset \citep{Faralli2015, AnilKumarTrikhaFattaneZarrinkalam}. \cite{AnilKumarTrikhaFattaneZarrinkalam} leverages frequent pattern mining techniques to identify topic sets. Here, a topic set consists of the topics frequently appear together in user profiles. Afterwards, the other topics in the topic sets that contain the topics in a user's profile are added into that user's profile as well. 

Take an example from \cite{AnilKumarTrikhaFattaneZarrinkalam}, a topic set identified via FPM might consist of two topics $z_1$ and $z_2$, where $z_1=\{\texttt{Mixtape, Hip\_hop\_music, Rapping, Kanye\_West, Jay-Z, Remix}\}$ and $z_2=\{\texttt{Lady\_Gaga, Song, Album, Concert, Canadia\_Hot\_100}\}$. $z_1$ refers to the topic about hip hop music produced by two American rappers \texttt{Jay-Z} and \texttt{Kanye\_West} while $z_2$ represents the topic about \texttt{Lady\_Gaga}'s concert in Canada. As these two topics frequently appear together in user interest profiles, the users who are interested in $z_1$ might be also interested in $z_2$ even $z_2$ is not in their primitive interests. In contrast to \cite{AnilKumarTrikhaFattaneZarrinkalam}, \cite{Faralli2015} did not directly enhance user interest profiles with other interests that occur together frequently, but used FPM for user classification and recommendation. It is worth noting that both \cite{Faralli2015} and  \cite{AnilKumarTrikhaFattaneZarrinkalam} used the FP-Growth algorithm \citep{Han2000} for frequent pattern mining in their studies.

\subsubsection{Temporal Dynamics of User Interests}
\label{Temporal Dynamics of User Interests}

User interests in OSNs can change over time, and many studies have been conducted in order to investigate the temporal dynamics of user interests in OSNs. For example, \cite{Jiang2015} showed that the similarity of current user interest profiles with the profiles at the beginning of the observation period of their dataset is the lowest while the similarity of current profiles with the ones built in the last month is the highest. Similarly, \cite{Abel2011g} showed that a user interest profile built in an earlier week differs more from the current profile compared to one built recently. In order to incorporate the temporal dynamics of user interests into user modeling strategies, there are mainly two types of approaches: (1) \emph{constraint-based} approaches, and (2) \emph{interest decay functions}. 


\textbf{Constraint-based approaches.} Constraint-based approaches extract user interest profiles based on specified constraints, e.g., using a \emph{temporal constraint} to build user interest profiles based on their tweets posted in the last two weeks or using an \emph{item constraint} to construct user profiles based on the last 100 tweets of the users. For example, \cite{Abel2011g} investigated several temporal constraints such as \emph{long-} and \emph{short-term}, and \emph{weekend} in their user modeling strategies on Twitter for a news recommender system. \emph{Long-term} profiles extract user interests from entire historical tweets of users while \emph{short-term} profiles extract user interests from tweets posted within the last two weeks. They showed that long-term entity-based profiles outperform short-term ones in the context of news recommendations. User interests can be different within different time frames such as during the week or on the weekends. The experimental results in \cite{Abel2011g} also showed that entity-based interest profiles  based on their tweets posted on weekends can outperform long-term profiles for recommending news on weekends. 

Some interests of users such as professional interests are stable while other interests such as the ones related to a certain event can be temporary. A user modeling strategy can apply temporal dynamics selectively to different information sources based on their characteristics. This type of strategy has been adopted in practical user modeling systems such as the one in Klout \citep{Spasojevic:2014:LLS:2623330.2623350}, in which a 90 day window is used for capturing the temporal dynamics of user interests for some temporal information sources, and an all-time window is used for more permanent sources such as professional interests. 

\cite{Nishioka:2016:PVT:2910896.2910898} compared both constraint-based approaches and interest decay functions for constructing user interest profiles on Twitter in the context of publication recommendations. Differing from the results in the domain of news \citep{Abel2011g}, results from \cite{Nishioka:2016:PVT:2910896.2910898} showed that a constraint-based approach constructing user interest profiles within a certain period performs better than using an interest decay function in the context of publication recommendations. 


\textbf{Interest decay functions.} Constraint-based approaches include interests which meet predefined constraints, and exclude other interests completely. Instead of constructing user interest profiles in a certain period (e.g., short-term), or based on temporal patterns (e.g., weekends), interest decay functions aim at including all the interests of a user but decaying old ones. The intuition behind those interest decay functions is that a higher weight should be given to recent interests than old ones. 

A popular type of interest decay function applies exponential decay to user interests. For example, the interest decay function from \cite{Orlandi2012} is defined as follows: 

\begin{equation}
	\label{eq:orlandi}
	x(t)=x_0 \cdot e^{-t/\beta}
\end{equation}
\\
Here, $x(t)$ is the decayed weight at time \emph{t}, and $x_0$ denotes the initial weight (at time $t=0$). This interest decay function also has an initial time window (7 days), and the interests in the time window are not discounted. The authors in  \cite{Orlandi2012} set $\beta =360days$ and $\beta =120days$ for their experiment, and showed that using $\beta =360days$ performs better than using $\beta =120days$ in terms of an evaluation based on a user study. We use \texttt{decay(Orlandi)} to denote this approach in this study. A similar decay function was used in \cite{Bhargava:2015:UMU:2678025.2701365} and \cite{Nishioka:2016:PVT:2910896.2910898}, where a weight for the last update was used instead of initial weight \citep{Bhargava:2015:UMU:2678025.2701365}. In \cite{OBanion2012}, the authors also used an exponential decay function: $x(t) = x_0 \cdot 0.9^d$ where \emph{d} is the difference in days between the current date and the date that a concept was mentioned. 

\cite{Abel2011d} also proposed a time-sensitive interest decay function, which is denoted by \texttt{decay(Abel)} in this survey. The weight of an entity \emph{e} with respect to a user \emph{u} at a specific time is measured as below.

\begin{equation}
	\label{eq:abel}
	w(e, time, T_{tweets, u, e}) = \sum_{t \in T_{tweets, u, e}} (1 - \frac{|time-time(t)|}{max_{time}-min_{time}})^d
\end{equation}
\\
where $T_{tweets, u, e}$ denotes the set of tweets mentioning \emph{e} that have been posted by \emph{u}. \emph{time}(\emph{t}) denotes the timestamp of a given tweet \emph{t}, and $max_{time}$ and $min_{time}$ denote the highest (youngest) and lowest (oldest) timestamp of a tweet in $T_{tweets, u, e}$. In addition, the parameter \emph{d} determines the influence of the temporal distance \citep[$d=4$ in][]{Abel2011d}. In contrast to the aforementioned exponential decay functions, this approach incorporates the age of an entity \emph{e} at the recommendation time, and the time span of \emph{e} with respect to \emph{u}.

In order to compare different interest decay functions in the context of user modeling in OSNs, \cite{piao2016exploring} investigated three interest decay functions for constructing user interest profiles on Twitter including \texttt{decay(Abel)} and \texttt{decay(Orlandi)}. The other one is a modified interest decay function from \cite{Ahmed2011}, which was used in advertisement recommendations on web portals (i.e., Yahoo!\footnote{\url{https://yahoo.com/}}). The modified interest decay function used in \cite{piao2016exploring} is defined as follows:

\begin{equation}
	\label{eq:ahmed}
	w_{ik}^t = \mu_{2week}w_{ik}^{t, week} + \mu_{2month}w_{ik}^{t, month}  + \mu_{all}w_{ik}^{t, all} 
\end{equation}
\\
where $\mu_{2week} = \mu$, $\mu_{2month} = \mu^2$ and $\mu_{all} = \mu^3$ where $\mu = e^{-1}$. This decay function combines three levels of abstractions where the decay of user interests in each abstraction is $\mu$ times the previous abstraction. We use \texttt{decay(Ahmed)} to denote this approach in this survey. \cite{piao2016exploring} conducted a comparative study of user interest profiles constructed based on the three aforementioned interest decay functions and the profiles based on \emph{short-} and \emph{long-term} periods. Those interest profiles were then evaluated in the context of URL recommendations. The results showed that using \texttt{decay(Ahmed)} and \texttt{decay(Orlandi)} have competitive performance in terms of URL recommendations, and perform better than using \texttt{decay(Abel)} as well as \emph{short-} and \emph{long-term} profiles which were constructed without any interest decay. In addition, the experimental results indicate that although the performance increases by giving a higher weight to recent user interests, it starts decreasing once the weight of recent interests is too high. That is, although applying the decay function to recent user interests increases the performance, we still need the old history in order to provide the best performance in the context of URL recommendations. 


Instead of considering the temporal dynamics of user interests with respect to individual users, global trends in an OSN can be incorporated into a user modeling strategy. In \cite{Gao:2011:ITU:2052138.2052335}, the authors combined user interests from tweets of a target user (user profiles) and of all users (trend profiles) for constructing user interest profiles. The TF weighting scheme is used for constructing user profiles. For trend profiles, they applied a time-sensitive TF$\cdot$IDF (t-TF$\cdot$IDF) weighting scheme to concepts:

\begin{equation}
	w_{t-TF \cdot IDF}(I_j, c) = w_{TF \cdot IDF}(I_j, c) \cdot (1-\hat{\sigma}(c))
\end{equation}
\\
where $w_{TF \cdot IDF}(I_j, c)$ denotes the TF$\cdot$IDF score of a concept \emph{c} in a given time interval $I_j$, and $\hat{\sigma}(c)$ denotes the normalized standard deviation of timestamps of tweets that refer to \emph{c}. \cite{Kanta2012} further incorporated location-aware trends into the trend-aware user modeling approach in \cite{Gao:2011:ITU:2052138.2052335} to improve the performance of inferred user interest profiles in the context of news recommendations.

\subsection{Summary and discussion}

This section reviewed a number of approaches for constructing and enhancing user interest profiles. Table \ref{construction} summarizes the approaches discussed in this section in terms of the three dimensions: (1) weighting schemes for constructing primitive interests, (2) approaches for incorporating the temporal dynamics of user interests, and (3) profile enhancement methods.

\renewcommand{\arraystretch}{1}
\afterpage{
\begin{landscape}
\begin{table}[]
\centering
\caption{Approaches for constructing and enhancing user interest profiles. Abbreviations are used as follows in this table: Prob. = probability, Int. Dec. = interest decay, Hie. = hierarchical knowledge, Gra. = graph-based knowledge, Col. = collective knowledge.}
\label{construction}
\begin{tabular}{|c|c|c|c|c|c|c|l|}
\hline
\multicolumn{2}{|c|}{\textbf{Weighting Scheme}} & \multicolumn{2}{c|}{\textbf{Temporal Dynamics}} & \multicolumn{3}{c|}{\textbf{Profile Enhancement}} & \multirow{2}{*}{\textbf{Examples}}
\\ \cline{1-7}
\textbf{\begin{tabular}[c]{@{}l@{}}Heuristics \end{tabular}} & \textbf{\begin{tabular}[c]{@{}l@{}}Prob. \\Inference \end{tabular}} & \textbf{\begin{tabular}[c]{@{}l@{}} Constraint\\-based \end{tabular}} & \textbf{\begin{tabular}[c]{@{}l@{}} Int. Dec.\\ Functions \end{tabular}} & \textbf{\begin{tabular}[c]{@{}l@{}} Hie. \end{tabular}} & \textbf{\begin{tabular}[c]{@{}l@{}} Gra. \end{tabular}} & \textbf{\begin{tabular}[c]{@{}l@{}} Col. \end{tabular}} &  \\ \hline

\checkmark            & & & & & & & \begin{tabular}[c]{@{}l@{}} \cite{Hannon2012},  \cite{Kapanipathi2011}, 
\\ \cite{Lim:2013:ICT:2491055.2491078}, \cite{Narducci2013},
\\ \cite{Abel2011e}, \cite{Bhattacharya:2014:IUI:2645710.2645765},
\\ \cite{Zarrinkalam2015a}, 
\\ \cite{Zarrinkalam2016}, \cite{Chen2010}, 
\\ \cite{GarciaEsparza:2013:CCT:2449396.2449402}, \cite{Ahn:2012:IUI:2457524.2457681}
\\  \cite{Vu:2013:IMU:2505515.2507883}, \cite{Phelan:2009:UTR:1639714.1639794},
\\  \cite{Karatay2015a},        \end{tabular}      \\ \hline

& \checkmark & & & & & & \begin{tabular}[c]{@{}l@{}}  \cite{Weng:2010:TFT:1718487.1718520}, \cite{Xu2011}     \end{tabular}      \\ \hline

\checkmark            & & \checkmark  &  & & & & \begin{tabular}[c]{@{}l@{}}   \cite{Abel2011g, Abel:2013:TUM:2540128.2540558}     \end{tabular}      \\ \hline

\checkmark            & & & \checkmark  & & & & \begin{tabular}[c]{@{}l@{}}   \cite{Gao:2011:ITU:2052138.2052335}, \cite{Kanta2012},
\\ \cite{Abel2011d}, \cite{OBanion2012}       \end{tabular}      \\ \hline

& \checkmark & \checkmark &   & & & & \begin{tabular}[c]{@{}l@{}}  \cite{Sang:2015:PFT:2806416.2806470}   \end{tabular}      \\ \hline

\checkmark            & & &  & \checkmark & & & \begin{tabular}[c]{@{}l@{}}  \cite{Bolting2015}    \end{tabular}      \\ \hline

\checkmark            & & &  & &  \checkmark & & \begin{tabular}[c]{@{}l@{}}   \cite{Lu2012}, 
\\ \cite{Abel2012a},  \cite{Penas2013},
\\ \cite{Piao2017, Piao2016b}       \end{tabular}      \\ \hline

\checkmark            & & &  & \checkmark &   & & \begin{tabular}[c]{@{}l@{}}   \cite{Kapanipathi2014},  \cite{Jipmo2017},
\\ \cite{Michelson2010}, 
\\ \cite{Kang2016}, \cite{Besel:2016:ISI:2851613.2851819, Besel:2016:QSI:3015297.3015298},  
\\ \cite{Faralli2017}, \cite{Nechaev},
\\ \cite{Jiang2015}, \cite{Siehndel:2012:TUP:2887379.2887395}        \end{tabular}      \\ \hline

\checkmark            & & &  & \checkmark & \checkmark  & & \begin{tabular}[c]{@{}l@{}}   \cite{Guangyuan2017}      \end{tabular}      \\ \hline

\checkmark            & & &  &  & \checkmark  & \checkmark & \begin{tabular}[c]{@{}l@{}}   \cite{Faralli2015}     \end{tabular}      \\ \hline

    &  \checkmark & \checkmark &  & &   & \checkmark  & \begin{tabular}[c]{@{}l@{}}   \cite{AnilKumarTrikhaFattaneZarrinkalam}     \end{tabular}      \\ \hline

\checkmark            & & &   \checkmark & \checkmark &   & & \begin{tabular}[c]{@{}l@{}} \cite{Bhargava:2015:UMU:2678025.2701365}    \end{tabular}      \\ \hline

\checkmark            & & \checkmark & & \checkmark &   & & \begin{tabular}[c]{@{}l@{}}   \cite{Nishioka:2015:ITU:2809563.2809601}, \cite{Budak2014}     \end{tabular}      \\ \hline

\checkmark            & & \checkmark &  & & \checkmark   & & \begin{tabular}[c]{@{}l@{}}  \cite{Spasojevic:2014:LLS:2623330.2623350}  \end{tabular}      \\ \hline

\checkmark            & &  & \checkmark & & \checkmark   & & \begin{tabular}[c]{@{}l@{}}   \cite{Orlandi2012}, \cite{piao2016exploring, Piao2016d}    \end{tabular}      \\ \hline

\checkmark            & & \checkmark & \checkmark & \checkmark &   & & \begin{tabular}[c]{@{}l@{}}   \cite{Nishioka:2016:PVT:2910896.2910898}    \end{tabular}      \\ \hline

\end{tabular}
\end{table}
\end{landscape}
}

As we can see from the table, many studies have incorporated the temporal dynamics of user interests in their user modeling strategies. Among interest decay functions, exponential decay functions such as \texttt{decay(Orlandi)} have been adopted widely. When incorporating the temporal dynamics of user interests, it is important to choose constraint-based approaches or interest decay functions based on the purpose of user modeling. For instance, when using inferred user interest profiles for recommending items such as news or URLs in OSNs, interest decay functions perform better than constraint-based approaches such as short- and long-term profiles \citep{piao2016exploring}. However, the results from \cite{Nishioka:2016:PVT:2910896.2910898} indicate that a constraint-based approach based on a certain period for profiling outperforms the one applying exponential decay for building user profiles in the context of a publication recommender system. One possible explanation is that user interests change differently with respect to different domains. For example, user interests should be adapted to their recent interests for news or URL recommendations, however, user interests with respect to research may not. 

\cite{Jiang2015} also pointed out that users have two types of interests; (1) \emph{stable interests} \citep[which they call primary interests in][]{Jiang2015}, and (2) secondary interests. The stable interests of a user are original preferences inherent to that user, such as programmers who like efficient algorithms or lawyers who like debate, etc. \citep{Jiang2015}. In contrast, secondary interests are temporary ones which closely follow hot topics or events in a specific timespan. This is in line with the user modeling strategy used in Klout \citep{Spasojevic:2014:LLS:2623330.2623350}, which applies a short-term window for capturing user interests that are temporary and uses a long-term window for more stable user interests. 

Different types of knowledge from various knowledge bases have been leveraged for enhancing the primitive interests of users. The diversity of KBs and the different structures of hierarchical KBs indicate the complexity of representing knowledge in KBs as well. Table \ref{topicTree} summarizes the differences between hierarchical KBs used in the literature. For instance, the constructed Wikipedia category taxonomy in \cite{Kapanipathi2014} consists of 15 levels with 802,194 categories while the topic hierarchy tree built by \cite{Jiang2015} consists of 5 levels with over 1,000 topics. The topic hierarchy tree used in Klout has 3 levels which consists of 15 main categories, around 700 sub-categories, and around 9,000 entities \citep{Spasojevic:2014:LLS:2623330.2623350}. A concept taxonomy built manually by referring to external websites such as news portals or Facebook Page categories has less complexity compared to a taxonomy based on KBs such as Wikipedia. For example, the category taxonomy built based on news portals \citep{Kang2016} has 8 main categories and 58 sub-categories. The one built based on Facebook and Yelp categories \citep{Bhargava:2015:UMU:2678025.2701365} also has 8 and 58 categories for the top-2 levels with an additional 137 categories in its third level. We can observe that the hierarchical knowledge bases used in practice or built based on taxonomies used in practice tend to have a small number of levels (2-5). Applying a spreading activation function, even the same one, to those different taxonomies might have different results. There is a lack of comparison of different hierarchical knowledge bases and their effect in the context of inferring user interest profiles.

Furthermore, although some studies investigated the comparison between using different KBs such as Wikipedia categories and the DBpedia graph, there was no comparative study on exploiting the Wikipedia entity graph \citep{Lu2012}, categories in other KBs such as ODP, and the DBpedia graph. In addition, despite the fact that different KBs might be useful in different domains \citep{Nguyen}, enhancing user interests based on other KBs such as Wikidata \citep{Vrandecic2014}, or BabelNet \citep{NavigliPonzetto:12aij} has not been fully explored.

\begin{table*}[!h]
\centering
\begin{tabular}{ |c|c|c|c| } 
\hline
\textbf{Study} &  \textbf{\# Levels} &  \textbf{\# Topics} &  \textbf{Details} \\ \hline
\cite{Kapanipathi2014} & 15 & 802,194 & N/A \\ \hline
\cite{Jiang2015} & 5 & $\sim$1,000 & N/A \\ \hline
\cite{Spasojevic:2014:LLS:2623330.2623350} & 3 & $\sim$1,0000 & 15 $\rightarrow$ $\sim$700 $\rightarrow$ $\sim$9,000 \\ \hline
\cite{Kang2016} & 2 & 66 & 8 $\rightarrow$ 58 \\ \hline
\cite{Bhargava:2015:UMU:2678025.2701365} & 3 & 203 & 8 $\rightarrow$ 58 $\rightarrow$ 137 \\ \hline
\end{tabular}
\caption{The structures of hierarchical knowledge bases for representing topics in different studies. The final column shows the number of concepts in each level of the corresponding hierarchical knowledge base.}
\label{topicTree}
\end{table*}



\section{Evaluation Approaches}

\subsection{Overview}
In this section, we describe evaluation approaches used for evaluating different user interest profiles that are generated by different user modeling strategies in the literature. User modeling is one of the main building blocks in many adaptive systems such as recommender systems. Many previous studies on the evaluation of adaptive systems suggested that it is important to evaluate different blocks separately in order to identify the problems in the adaptive systems \citep{Paramythis2010, Brusilovsky2001}. \cite{Gena2007} provided a list of methods for evaluating adaptive systems, where some of them can be used for evaluating the quality of user modeling component as well. These evaluation methods include (1) \emph{questionnaires}, (2) \emph{interviews}, and (3) \emph{logging use}.

\textbf{Questionnaires.} Questionnaires consist of pre-defined questions, which can be in different styles such as scalar or multi-choice, and ranked \citep{Gena2007}. In our context, this approach can be used for collecting users' explicit feedback about their interest profiles for evaluation. To this end, this approach requires recruiting users for the experiment of building user interest profiles with their OSN accounts. At the end of the experiment, these users can provide feedback on user interest profiles constructed by different user modeling strategies.

\textbf{Interviews.} The second approach is used to collect users' opinions and experiences, preferences and behavior motivations \citep{Gena2007} with respect to adaptive systems. Interviews can be used after building users' interest profiles to gather their opinion such as satisfaction and accuracy about the inferred user interest profiles. Compared to questionnaires, interviews are more flexible but more difficult to be administered. Therefore, this method has not been exploited for evaluating user modeling strategies in the literature. 

\textbf{Extrinsic evaluation (Logging use).} This approach uses the actions of users in the context of adaptive systems for evaluation, e.g., whether a user liked a recommend item in a recommender system. This can be considered an extrinsic way of evaluating user interest profiles in terms of the performance of applications where these profiles are applied. For example, one common approach is using constructed user interest profiles as an input to a recommender system, and adopting some well-established evaluation metrics of recommender systems for measuring the quality of user interest profiles indirectly. Manual analysis is sometimes used together with other evaluation approaches. In this case, the authors present some examples of user interest profiles built for several users (e.g., some representative users on Twitter such as \emph{Barack Obama}), and discuss the quality of profiles with respect to these users.

\subsection{Review}

\subsubsection{Evaluation based on Questionnaires}

A common approach for evaluating constructed user interest profiles is based on a user study with questionnaires. For example, \cite{Narducci2013} evaluated user interest profiles built for 51 users from Facebook and Twitter based on their feedback on two aspects: \emph{transparency} and \emph{serendipity} using a 6-point discrete rating scale. The first aspect aims to evaluate to what extent the keywords in the profile reflect personal interests, and the second one aims to measure to what extent the profile contains unexpected interesting topics. Similarly, \cite{Kapanipathi2014} recruited 37 users and built category-based user interest profiles based on their tweets on Twitter. Afterwards, the 37 users provided explicit feedback, e.g., Yes/Maybe/No with respect to the categories in those profiles. Similar approaches have been used in \cite{Bhattacharya:2014:IUI:2645710.2645765}, \cite{Besel:2016:ISI:2851613.2851819, Besel:2016:QSI:3015297.3015298}, \cite{Budak2014}, and \cite{Orlandi2012}. However, instead of recruiting volunteers for an experiment, the authors in \cite{Budak2014} first inferred user interest profiles for 500 randomly chosen users on Twitter, and emailed them using the email addresses in their profiles to get feedback about their inferred interests. Instead of using the feedback from target users for inferred user interest profiles, \cite{Kang2016} and \cite{Michelson2010} labeled user interests themselves or used recruited annotators. 

Explicit feedback can be obtained in a system which has user interest profiles that can be modifed by users. For example, \cite{GarciaEsparza:2013:CCT:2449396.2449402} implemented a stream filtering system where users are represented based on 18 defined categories such as \texttt{Music} and \texttt{Sports}. For evaluation, the authors asked each participant to give explicit feedback on their profiles by deleting or adding categories that they felt were incorrect or missing. 

In contrast to obtaining explicit feedback on inferred user interest profiles, a user study can be conducted on the performance of a specific application where those inferred user interest profiles play an important role. For example, \cite{Chen2010} conducted a user study with respect to a URL recommender system on Twitter, which is based on the inferred user interest profiles. Therefore, instead of directly giving feedback on the constructed user interest profiles, the users participating in the study were given URL recommendations, and they marked each URL as one of their interests or not. Similarly, \cite{Nishioka:2016:PVT:2910896.2910898} obtained explicit feedback from users on publication recommendations based on their interest profiles. These user studies can also be considered as extrinsic evaluation, which we will discuss in the next section, as they are not evaluating user interest profiles directly.
\\\\
\textbf{Pros and cons.} Evaluation approaches based on the explicit feedback of profiled users with respect to their interest profiles would arguably be the most direct and accurate way for evaluating those profiles. However, this also requires recruiting volunteers and imposes an extra burden for users, and therefore limits the number of participants for evaluation \citep[e.g., 37 users were recruited for evaluation in][]{Kapanipathi2014}.

\subsubsection{Extrinsic Evaluation}

To evaluate the quality of inferred user interest profiles without imposing an extra burden on users, offline evaluation in terms of the performance of a specific application has been used. In this case, user interest profiles are used as an input to an application such as a news recommender system where these profiles play an important role. Afterwards, different profiles created by different user modeling strategies are compared in terms of the recommendation performance using each profile. The recommendation performance can be evaluated by well-established evaluation metrics for recommender systems such as \emph{mean reciprocal rank} (MRR) which denotes at which rank the first item relevant to the user occurs on average, \emph{success at rank N} (S@N), which stands for the mean probability that a relevant item occurs within the top-N recommendations, and well-known \emph{precision} and \emph{recall}. For a complete list of evaluation metrics and their details we refer the reader to \cite{Bellogn} and \cite{Herlocker:2004:ECF:963770.963772} respectively.

For instance,  \cite{Abel2011g} evaluated three different user modeling strategies in terms of S@N and MRR in the context of news recommendations, and \cite{Spasojevic:2014:LLS:2623330.2623350} evaluated their user modeling strategy in terms of precision and recall in the context of topic recommendations on Klout. Similarly, \cite{Sang:2015:PFT:2806416.2806470} also evaluated user interest profiles in terms of news recommendations in addition to tweet recommendations. \cite{piao2016exploring, Guangyuan2017, Piao2017, Piao2016b, Piao2016d} evaluated different user modeling strategies in the context of URL recommendations on Twitter where the set of ground truth URLs is those shared by users on Twitter in the last two weeks. In \cite{Faralli2015}, the authors evaluated user interest profiles in terms of user classifications and recommendations. For the classification task, the user interest profiles were used for classifying each user to the appropriate label, e.g., Starbucks fan. For the recommendation task, the authors evaluated the performance of leveraging different hierarchical levels of interests with respect to interest recommendations using itemset mining. 

In contrast to previous studies which have focused on inferring user interest profiles, \cite{Nechaev} focused on users' privacy and evaluated different followee-suggestion strategies for concealing user interests which can be inferred from users' activities in OSNs based on state-of-the-art user modeling strategies.
\\\\
\textbf{Pros and cons.} Extrinsic evaluation provides an offline setting for evaluating inferred user interest profiles. Therefore, it facilitates the evaluation process of different user modeling strategies as these strategies are evaluated based on a collected dataset (or logs). However, this approach does not directly evaluate the inferred user interest profiles, and lacks the opinions of users with respect to the inferred interest profiles.
\\\\
There are other evaluation approaches used in some studies besides the aforementioned two methods. For example, \cite{Abel2011e} compared the number of distinct entities and topics in user interest profiles for evaluating news-based enrichment of their tweets. In \cite{Faralli2017}, the authors run two experiments to evaluate their approach of building interest taxonomies. First, they compared their approach against other approaches proposed for constructing user interest taxonomies using other gold standard taxonomies. Second, they provided samples of generated user interest profiles, and compared inferred Wikipedia categories with respect to several users based on different user modeling strategies. Similarly, \cite{Xu2011} evaluated their topic modeling approach by comparing it against other topic modeling methods in terms of \emph{perplexity}, and then discussed some user interest profiles produced by different approaches. User interest profiles have also been used for specific applications such as followee, tweet, and news recommendations \citep{Weng:2010:TFT:1718487.1718520, Chen2012b, Hong:2013:CMM:2433396.2433467,Phelan:2009:UTR:1639714.1639794}, where user modeling strategies were not evaluated or compared to other alternatives.

\subsection{Summary and discussion}

\begin{table}[!b]
\centering
\caption{Evaluation approaches for constructed user interest profiles.}
\label{evaluation}
\begin{tabular}{|c|c|l|}
\hline
\textbf{\begin{tabular}[c]{@{}l@{}}Questionnaires\end{tabular}}  &
\textbf{\begin{tabular}[c]{@{}l@{}}Extrinsic \\Evaluation\end{tabular}} 
 & \textbf{Examples} \\ \hline
\checkmark &       & \begin{tabular}[c]{@{}l@{}} \cite{Kapanipathi2014}, \cite{Kang2016},  \\  \cite{Michelson2010}, \cite{Budak2014}, \\ \cite{Bhattacharya:2014:IUI:2645710.2645765}, \cite{Besel:2016:ISI:2851613.2851819, Besel:2016:QSI:3015297.3015298}, \\ \cite{Orlandi2012}, \cite{Narducci2013}, \\ \cite{Bhargava:2015:UMU:2678025.2701365}, \cite{GarciaEsparza:2013:CCT:2449396.2449402}, \\ \cite{Vu:2013:IMU:2505515.2507883}, \cite{Ahn:2012:IUI:2457524.2457681}, \\ \cite{Chen2010}, \cite{Nishioka:2016:PVT:2910896.2910898}    \end{tabular}    \\ \hline
& \checkmark &  \begin{tabular}[c]{@{}l@{}} \cite{Abel2011d, Abel2011g, Abel2012a, Abel2011e}, \cite{Chen2010}, \\ \cite{Zarrinkalam2015a}, \cite{Sang:2015:PFT:2806416.2806470},  \\ \cite{Kanta2012}, \cite{OBanion2012},  \\ \cite{piao2016exploring, Guangyuan2017, Piao2017, Piao2016b, Piao2016d},
\\ \cite{Lu2012}, \cite{Sang:2015:PFT:2806416.2806470}, \cite{Gao:2011:ITU:2052138.2052335},
\\ \cite{Karatay2015a}, \cite{AnilKumarTrikhaFattaneZarrinkalam}, 
\\ \cite{Nishioka:2015:ITU:2809563.2809601}, \cite{Bolting2015}, 
\\ \cite{Zarrinkalam2016}, \cite{Ahn:2012:IUI:2457524.2457681}, \\ \cite{Spasojevic:2014:LLS:2623330.2623350}, \cite{Jipmo2017},
\\ \cite{Faralli2015}, \cite{Nechaev}    \end{tabular} \\ \hline
\end{tabular}
\end{table}

In this section, we reviewed different evaluation approaches that have been used in the literature for evaluating constructed user interest profiles. Table \ref{evaluation} provides a summary of previous studies in terms of evaluation methods.

Evaluating user interest profiles based on a user study is important for understanding different aspects of user interests, e.g., abstraction levels of user interests. For example, \cite{Orlandi:2013:CCI:2568488.2568810} studied the specificity of user interests and evaluated it based on a user study, which showed that users prefer to give a higher score over non-specific entities. However, the extra effort of recruiting users and gaining feedback from them is time consuming, and limits the scale of users for evaluation. The evaluation in terms of the performance of a specific application has the advantage of its offline setting and using a relatively larger number of users compared to a user study. Both evaluation approaches can be used in an appropriate way for designing and evaluating user modeling strategies. For example, based on a user study on the specificity of user interests \citep{Orlandi:2013:CCI:2568488.2568810}, we can design ways to incorporate the feedback from users' preferences regarding non-specific entities into a user modeling strategy, and evaluate the strategy at a large scale in offline settings based on a collected dataset such as the one from Twitter.

One of the challenges of the offline evaluation in terms of the performance of a specific application is the lack of benchmarks that are freely available \citep{Faralli2015}. Despite the openness of some microblogging services such as Twitter, it is time consuming to collect all data used in different user modeling approaches, e.g., tweets, list memberships, biographies of followees/followers in addition to the information about users. In addition, different datasets with different user sizes might produce different results even using the same user modeling strategies for comparison. It is also important to evaluate different user interest profiles in the context of different applications beyond a specific one. For example, in \cite{Manrique:2017:SDA:3106426.3109440}, the authors showed that user interest profiles based on different user modeling strategies perform differently in the context of recommending articles based only on titles, abstracts, and full texts. Although the study \citep{Manrique:2017:SDA:3106426.3109440} is in the context of research article recommendations, it is highly likely that different user interest profiles from microblogging services will have different levels of performance based on the applications in which these profiles are applied.




\section{Conclusions and Future Directions}

In previous sections, we reviewed the state-of-the-art approaches used in different user modeling stages for inferring user interest profiles, which is beneficial both for researchers who are interested in user modeling in the social networks domain as well as those researchers in some other domains. It is also useful for third-party application providers who aim to utilize user interest profiles via social login functionalities in terms of providing personalized services for their users. 

In this final section, we conclude this paper in Section \ref{conclusions} with respect to the four dimensions of inferring user interest profiles: (1) data collection, (2) representations of user interest profiles, (3) construction and enhancement of user interest profiles, and (4) the evaluation of the constructed profiles. In Section \ref{fd}, we first review what progress has been made to date since \cite{Abdel-Hafez2013}, and then outline some opportunities and challenges for inferring user interests on microblogging social networks which we envision can inspire future directions in this research field.

\subsection{Conclusions}\label{conclusions}
To sum up, user activities such as the tweets posted by users are the most widely used information source for inferring user interests. However, many recent studies have started exploring other information sources such as the social networks of users as an alternative to user activities as the passive usage of OSNs is on the rise. Regarding the representations of user interest profiles, a clear tendency of leveraging concepts such as DBpedia entities or categories can be observed given their advantages of using background knowledge about those concepts from a KB. In addition to leveraging the hierarchical or graph-based knowledge of a KB for enriching user interests, several recent studies also have shown the effectiveness of leveraging collective knowledge for enriching user interest profiles \citep{Faralli2015, AnilKumarTrikhaFattaneZarrinkalam}. With respect to incorporating the temporal dynamics of user interests, there is no single best method for inferring user interests with different purposes. Instead, one should choose constraint-based or interest decay functions based on the application needs, and the characteristics of items. For evaluating user interest profiles, both questionnaires and extrinsic evaluation strategies have been adopted at comparable levels of popularity.

\subsection{Future Directions}\label{fd}

In \cite{Abdel-Hafez2013}, the authors proposed three future directions with respect to user modeling in OSNs, which requires (1) more dynamicity, (2) more enrichment, and (3) more comprehensiveness. On the one hand, we observe that there have been many efforts towards the second direction. These efforts include leveraging the collective knowledge powered by all users \citep{Faralli2015, AnilKumarTrikhaFattaneZarrinkalam} for enriching the interest profiles of each user, and the comparison between different KBs for enriching user interests \citep{Guangyuan2017}. On the other hand, the first and third directions proposed by \cite{Abdel-Hafez2013} have not made much progress. For example, \cite{Abdel-Hafez2013} proposed incorporating more dynamicity with respect to user interest profiles with some assumptions such as different topics might decay with different speed, and the interest weights of each user can have different weights in different context. On top of the directions proposed by \cite{Abdel-Hafez2013} and the recent studies we reviewed in this paper, we further proposed several future directions which are related to:

\begin{itemize}
	\item mining user interests;
	\item multi-faceted user interests;
	\item comprehensive user modeling;
	\item evaluation of user modeling strategies.
\end{itemize}

\textbf{Mining user interests.} To better infer user interests, researchers have proposed various approaches such as enriching short content, filtering noise in UGC, and exploring social networks. Many studies have adopted traditional weighting schemes from information retrieval such as TF or TF$\cdot$IDF to somehow filter the noise in UGC for mining user interests. However, some studies have shown that incorporating some special characteristics of the services (e.g., temporal dynamics, short content) into the design of a weighting scheme can improve the quality of user interest profiles. For example, TI-TextRank which combines TF$\cdot$IDF and TextRank performs better than either of them on their own as a weighting scheme for user modeling on Twitter. In this regard, more weighting schemes adapted towards microblogging services should be investigated, e.g., combining different weighting schemes used in the literature. Furthermore, mining interest-related items from data sources such as posts \citep[e.g.,][]{Xu2011} can be useful as microblogging services have multiple usages such as information seeking, sharing and social networking \citep{Java2007}.

In addition, more sophisticated approaches for understanding the semantics of UGC are required. For example, for those approaches  that rely on extracted entities for inferring user interest profiles, extracting entities from microblogs is a fundamental step which is challenging by itself. Only a few studies have considered the uncertainty (confidence) of the extracted entities, which we think might impact the overall quality of the primitive interests of users as well as the enhanced ones. Moreover, most approaches have extracted explicitly mentioned entities based on NLP APIs such as Tag.Me\footnote{\url{https://tagme.d4science.org/tagme/}}, Aylien\footnote{\url{https://aylien.com/}}, OpenCalais, etc. However, there can be many entities implicitly mentioned in tweets. In \cite{Perera2016}, the authors showed that over 20\% of mentions of movies are implicit references, e.g., a tweet referring the movie \emph{Gravity} - ``ISRO sends probe to Mars for less money than it takes Hollywood to make a movie about it''. It shows that advanced methods for extracting entities, such as the one proposed in \cite{Perera2016}, have great potential to improve the quality of user modeling. Also, considering the context of a microblog might be useful when extracting entities instead of just considering the single microblog of a user. The context might refer to some previous microblogs posted by the user, or other microblogs with the same hashtag in the microblogging service. For example, \cite{Shen:2013:LNE:2487575.2487686} showed that the quality of entity extraction can be improved by incorporating user interests as contextual information. Furthermore, promising results from recent studies \citep{Faralli2015, AnilKumarTrikhaFattaneZarrinkalam} indicate that leveraging collective knowledge via frequent pattern mining approaches is also effective in inferring implicit user interests.

\textbf{Multi-faceted user interests.} There exists various aspects/views of users based on different dimensions of user modeling such as the data source, representation level, and temporal dynamics of user interests. Although many studies represent an individual user using a single user interest profile, we believe that multi-faceted user interest profiles should be given more attention as some previous studies have also shown their efficiency compared to a single model. It is not necessary to maintain several user interest profiles for a single user, but a single model can also be built with relevant information from different aspects, and a view/aspect made for the user based on the information needs for different applications. GeniUS \citep{Gao2012d} is a good example in this regard, which is a user modeling library that stores concept-based user interest profiles using the RDF\footnote{\url{https://www.w3.org/RDF/}} format (a W3C recommendation) with widely used ontologies such FOAF \citep{Brickley2012}, SIOC\footnote{\url{http://sioc-project.org/}}, and WI\footnote{\url{http://smiy.sourceforge.net/wi/spec/weightedinterests.html}}. In GeniUS, user interest profiles are represented as DBpedia entities and enriched by background knowledge such as the type (domain) of an entity from DBpedia. Therefore, the constructed profile is flexible enough to retrieve its sub-profiles with respect to specific domains (e.g., \texttt{Music}), which is useful for recommending domain-specific items. The idea is that, for example, we only need your music-related interest profile in the context of music recommendations. The results in \cite{Gao2012d} indicate that domain-specific profiles clearly outperform the whole user profiles for domain-specific tweet recommendations in terms of six different domains. Although GeniUS only considers different views of users in terms of topical domains, the same idea can be extended to other views. For instance, different user profiles can be extracted dynamically with different approaches for incorporating temporal dynamics, e.g., retrieving short-term profiles for recommending tweets during an event, which might be more useful compared to using long-term profiles. Also, multiple user interest profiles in terms of representation level using different interest formats have been used in other domains such as personal assistants \citep{Guha2015}, which can be useful for user modeling in microblogging services as well. In \cite{Guha2015}, several user interest profiles based on different representations such as keywords and Freebase entities were constructed. 

\textbf{Comprehensive user modeling.} In the previous survey on user modeling \citep{Abdel-Hafez2013}, the authors also suggested that more comprehensive user modeling strategies should be investigated by considering different dimensions of user modeling together. Many of the previous studies have ignored some of the dimensions such as temporal dynamics \citep[e.g.,][]{Phelan:2009:UTR:1639714.1639794}. Investigating the synergistic effect of different dimensions is important for developing better user modeling strategies, which is crucial for the performance of applications. To this end, several research questions should be answered such as ``which combinations of different approaches in each dimension can provide the best user interest profiles'' or ``does a dimension really matter in the context of the combination for providing the best performance?''. For example, \cite{Piao2016d} showed that a rich representation of user interests (using WordNet synsets and DBpedia entities) and enriching short content with the text of embedded URLs are the most important factors followed by temporal dynamics in the context of URL recommendations on Twitter. However, enhancing user interest profiles has little effect when we have a rich representation or enriched content of microblogs. Similar results have been observed in the context of inferring research interests of users based on their publications \citep{Manrique:2017:SDA:3106426.3109440}. The results in \cite{Manrique:2017:SDA:3106426.3109440} indicate that enhancing primitive interests can improve the performance when only short texts (e.g., titles) are available but not in the case when longer texts (e.g., full texts of publications) are available. We believe that these studies are good starting points for some future works, e.g., using different user interest profiles for different data sources instead of using a single representation of an individual user for the combination.

In addition, other user modeling dimensions which have been proposed in other domains can be considered in the social media domain as well. For example, a \emph{scrutable} user model proposed in the context of teaching, which aims to let users have the right and possibility to have access to and control their user profiles \citep{Holden1999, Carmagnola2011, Kay2006}, can be a promising dimension to be incorporated into user modeling strategies in OSNs and merits further investigation and evaluation.

\textbf{Evaluation of user modeling strategies.} As we mentioned in Section 5.3, the lack of common benchmarks and datasets hinders comparison with other approaches, which ends up with several studies directly comparing to results reported in previous studies \citep{Faralli2015}. This does not reflect a correct comparison due to the difference of datasets in terms of platforms as well as user sizes. However, it is also challenging due to the regulations of microblogging services such as Twitter\footnote{Twitter restrict developers from sharing the content of tweets, see \url{https://developer.twitter.com/en/developer-terms/agreement-and-policy}.}, and the differences in data sources used in each study. Another possible direction is providing all proposed approaches as user modeling libraries that are publicly available, in the same way as GeniUS and TUMS\footnote{Both GeniUS and TUMS are available at \url{http://www.wis.ewi.tudelft.nl/tweetum/}}, so that other researchers can easily reimplement the approaches proposed in previous studies for comparison.

It is also important to evaluate inferred user interest profiles in terms of multiple tasks or different settings to understand the strengths and weaknesses of different user interest profiles. For instance, \cite{Nishioka:2015:ITU:2809563.2809601} showed that considering the temporal dynamics of user interests has a positive influence on a computer science dataset but not on a medicine dataset.  \cite{Manrique:2017:SDA:3106426.3109440} showed that different user modeling strategies work differently for different types of texts that are available in the context of research article recommendations. In this regard, evaluating the performance of different user modeling strategies based on different datasets or settings can provide a clear understanding of when to use what types of user profiles, which is important for researchers in different domains as well as third-party application providers with different types of content to be personalized. A recent work by \cite{Tommasso2018} provides a user interests dataset which is useful in this context. It includes half million Twitter users with an average of 90 multi-domain preferences per user on music, books, etc., where those preferences are extracted from multiple platforms based on the messages of those Twitter users who also use Spotify\footnote{\url{https://www.spotify.com}}, Goodreads\footnote{\url{https://www.goodreads.com/}}, etc.

Finally, previous studies have adopted accuracy and ranking metrics such as precision, recall, and MRR for the extrinsic evaluation of inferred user interest profiles. However, non-accuracy metrics such as serendipity, novelty, and diversity have received increasing attention in recommender systems \citep{Bellogn, Kaminskas:2016:DSN:3028254.2926720}. Therefore, it is worth investigating the effect of different user modeling strategies and their inferred interest profiles in the context of recommender systems in terms of those non-accuracy metrics.

\begin{acknowledgements}
This publication has emanated from research conducted with the financial support of Science Foundation Ireland (SFI) under Grant Number SFI/12/RC/2289 (Insight Centre for Data Analytics). Thanks for the anonymous reviewers and the editor for their constructive feedback to improve this work.
\end{acknowledgements}

\bibliographystyle{spbasic}      
\bibliography{library}   

\section*{Author Biographies}
\begin{description}
    \item \textbf{Guangyuan Piao} (\url{https://parklize.github.io}) is a Ph.D. student at the Insight Centre for Data Analytics (formerly DERI) at the National University of Ireland Galway. He received his B.Sc. in Computer Science from Jilin University, China, and received his M.Eng. degree in Information and Industrial Engineering from Yonsei University, South Korea. His main research interests include User Modeling, Recommender Systems, and Knowledge Graph. His current research focuses on semantics-aware user modeling and recommender systems leveraging knowledge graphs and latent semantics.
\\
    \item \textbf{John G. Breslin} (\url{www.johnbreslin.com}) is a Senior Lecturer in Electrical and Electronic Engineering at the College of Science and Engineering at the National University of Ireland Galway, where he is Director of the TechInnovate / AgInnovate programmes. John has taught electronic engineering, computer science, innovation and entrepreneurship topics during the past two decades. He is also a Co-Principal Investigator at the Insight Centre for Data Analytics, and a Funded Investigator at Confirm Smart Manufacturing and VistaMilk. He has written 190 peer-reviewed academic publications (h-index of 37, 5500 citations, best paper awards from DL4KGS, SEMANTiCS, ICEGOV, ESWC, PELS), and co-authored the books "The Social Semantic Web" and "Social Semantic Web Mining". He co-created the SIOC framework, implemented in hundreds of applications (by Yahoo, Boeing, Vodafone, etc.) on at least 65,000 websites with 35 million data instances. 
\end{description}

\clearpage
\begin{appendices}
\section{The List of Surveyed Works}
\label{appendix:works}

\subsection{Search Strategy}

In order to draw up a list of search terms, the basic terms are extracted from primary articles are retrieved. After that, other search terms are obtained iteratively based on the keywords that were used interchangeably within the retrieved articles. Overall, the final list of terms used for searching articles is presented in Table \ref{terms}. These search terms (ST) are used for constructing sophisticated search strings. For example, the search string can be constructed as ST1 AND ST3 while ST1 is a compound term from Term1 and Term2 (e.g., inferring user interests). Initial searches with these search terms for titles and abstracts from electronic databases can obtain many relevant articles but may not be sufficient \citep{Kitchenham2004}. In this regard, additional article candidates are obtained by checking the reference list from primary studies that are relevant, and searching relevant journals and conference proceedings. \cite{Abdel-Hafez2013} provided a review of user modeling in social media websites in 2013, which includes some approaches with respect to inferring user interests in the context of microblogging social networks. In addition to those approaches mentioned in \cite{Abdel-Hafez2013}, we also review recent user modeling approaches for inferring user interests.

\begin{table}[!h]
\centering
\begin{tabular}{ |c|c|c| } 
 \hline
  & Term1 & Term2 \\ \hline
 ST1 & inferring, modeling, predicting & (user) interests \\ \hline
 ST2 & user (interest) & modeling, profiling, detection \\ \hline
 ST3 & \multicolumn{2}{c|}{social, online, twitter, microblogging} \\ 
 \hline
\end{tabular}
\caption{Search terms used in the search strategy of this survey.}
\label{terms}
\end{table}

\subsection{Selection Criteria}
In order to assess and select relevant articles from primary studies, inclusion and exclusion criteria should be defined based on the research questions \citep{Kitchenham2004}. The inclusion criteria are as follows:
\begin{enumerate}
	\item Published in English from 2004.
	\item Studies on microblogging social networks.
	\item Focus on user modeling strategies for inferring user interest profiles.
\end{enumerate}
On the other hand, exclusion criteria can be defined as follows:
\begin{enumerate}
	\item Studies that were not peer-reviewed or published.
	\item Studies related to user modeling but not focus on microblogging social networks.
	\item Studies related to user modeling, but not focus on inferring user interests.
\end{enumerate}

Finally, inclusion or exclusion decisions are made for the fully obtained articles and those papers that only meet our criteria are selected. As a result, 51 articles are selected in this survey. These articles are distributed from 2010 to 2018, and the majority of them were published in conferences or workshops such as WI, UMAP, CIKM, and ECIR.

\subsection{Surveyed Studies}
The surveyed 51 works are retrieved from different journals, conferences, and workshops, mainly in the user modeling, recommender systems, and Web related fields as follows:
\begin{enumerate}
	\item Journals
	 	\begin{itemize}
			\item ACM SIGAPP Applied Computing Review: \cite{Besel:2016:QSI:3015297.3015298}
			\item Web Semantics: Science, Services and Agents on the World Wide Web: \cite{Faralli2017}
			\item Social Network Analysis and Mining: \cite{Faralli2015}
			\item Information Systems: \cite{Kang2016}
			\item Procedia Computer Science: \cite{Jiang2015}
		\end{itemize}
	\item Conference proceedings
		\begin{itemize}
			\item \textbf{WI} (IEEE/WIC/ACM International Conference on Web Intelligence and Intelligent Agent Technology): \cite{Zarrinkalam2015a, Xu2011, Gao:2011:ITU:2052138.2052335,Penas2013, Ahn:2012:IUI:2457524.2457681}
			\item \textbf{UMAP} (Conference on User Modeling Adaptation and Personalization): \cite{ Abel2011g, Hannon2012, Narducci2013}
			\item \textbf{CIKM} (ACM International Conference on Information and Knowledge Management): \cite{Vu:2013:IMU:2505515.2507883, Piao2016b, Sang:2015:PFT:2806416.2806470}
			\item \textbf{ECIR} (European Conference on Information Retrieval): \cite{Zarrinkalam2016, Guangyuan2017, AnilKumarTrikhaFattaneZarrinkalam}
			\item \textbf{ISWC} (International Conference on Semantic Web): \cite{Siehndel:2012:TUP:2887379.2887395,Abel2011e}
			\item \textbf{IUI} (International Conference on Intelligent User Interfaces): \cite{Bhargava:2015:UMU:2678025.2701365, GarciaEsparza:2013:CCT:2449396.2449402}
			\item \textbf{RecSys} (ACM Conference on Recommender Systems): \cite{Bhattacharya:2014:IUI:2645710.2645765, Phelan:2009:UTR:1639714.1639794}
			\item \textbf{SEMANTiCS} (International Conference on Semantic Systems): \cite{piao2016exploring, Orlandi2012}
			\item \textbf{HT} (ACM Conference on Hypertext and Social Media): \cite{Piao2017}
			\item \textbf{SIGIR} (International ACM Conference on Research and Development in Information Retrieval): \cite{Chen2010}
			\item \textbf{AAAI} (AAAI Conference on Artificial Intelligence): \cite{Lu2012}
			\item \textbf{KDD} (Knowledge Discovery and Data Mining): \cite{Spasojevic:2014:LLS:2623330.2623350}
			\item \textbf{IJCAI} (International Joint Conference on Artificial Intelligence): \cite{Abel:2013:TUM:2540128.2540558}
			\item \textbf{ICWE} (International Conference on Web Engineering): \cite{Abel2012a}
			\item \textbf{WebSci} (International Web Science Conference): \cite{Abel2011d}
			\item \textbf{ESWC} (Extended Conference on Semantic Web): \cite{Kapanipathi2014}
			\item \textbf{EKAW} (International Conference on Knowledge Engineering and Knowledge Management): \cite{Piao2016d}
			\item \textbf{ICSC} (IEEE International Conference on Semantic Computing): \cite{Bolting2015}
			\item \textbf{SAC} (ACM Symposium on Applied Computing): \cite{Besel:2016:ISI:2851613.2851819}
			\item \textbf{WSDM} (ACM International Conference on Web Search and Data Mining): \cite{Weng:2010:TFT:1718487.1718520}
			\item \textbf{JCDL} (Joint Conference on Digital Libraries): \cite{Nishioka:2016:PVT:2910896.2910898}
			\item \textbf{i-KNOW} (International Conference on Knowledge Technologies and Data-driven Business): \cite{Nishioka:2015:ITU:2809563.2809601}
			\item \textbf{SPIM} (International Conference on Semantic Personalized Information Management: Retrieval and Recommendation): \cite{Kapanipathi2011}
			\item \textbf{OpenSym} (International Symposium on Open Collaboration): \cite{Lim:2013:ICT:2491055.2491078}
			\item \textbf{ADMA} (Advanced Data Mining and Applications): \cite{Jipmo2017}
		\end{itemize}
	\item Workshop proceddings
	\begin{itemize}
			\item \textbf{AND} (Workshop on Analytics for Noisy Unstructured Text Data): \cite{Michelson2010}
			\item \textbf{Micropost} (Workshop on Making Sense of Microposts): \cite{Karatay2015a}
			\item \textbf{SMAP} (Workshop on Semantic and Social Media Adaptation and Personalization): \cite{Kanta2012}
			\item \textbf{RSWeb} (Workshop on Recommender Systems and the Social Web): \cite{OBanion2012}
			\item \textbf{BlackMirror} (Workshop on Re-coding Black Mirror): \cite{Nechaev}
		\end{itemize}
	\item Others
	\begin{itemize}
			\item Tech Report: \cite{Budak2014}.
		\end{itemize}
\end{enumerate}
\end{appendices}

\end{document}